\documentclass[APS,reprint,superscriptaddress]{revtex4-1}

\usepackage{amsmath}    
\usepackage{graphicx}   
\usepackage{color}      
\usepackage{subfigure}  
\usepackage{hyperref}   
\usepackage{todonotes}
\usepackage[latin1]{inputenc}

\begin{document}

\title{Electronic properties of Mn-Phthalocyanine - C$_{60}$ bulk heterojunctions: combining photoemission and electron energy-loss spectroscopy}
\author{Friedrich Roth}
\affiliation{Center for Free-Electron Laser Science / DESY, Notkestra\ss e 85, D-22607 Hamburg, Germany}
\author{Melanie Herzig }
\affiliation{FW Dresden, PO Box 270116, D-01171 Dresden, Germany }
\author{Cosmin Lupulescu}
\affiliation{Inst. of Optics and Atomic Physics, TU Berlin, Stra\ss e des 17. Juni 135, D-10623 Berlin, Germany}

\author{Erik Darlatt}
\affiliation{Physikalisch-Technische Bundesanstalt (PTB), Abbestra\ss e 2-12, D-10587 Berlin, Germany}
\author{Alexander Gottwald}
\affiliation{Physikalisch-Technische Bundesanstalt (PTB), Abbestra\ss e 2-12, D-10587 Berlin, Germany}
\author{Martin Knupfer}
\affiliation{FW Dresden, PO Box 270116, D-01171 Dresden, Germany }
\author{Wolfgang Eberhardt}
\affiliation{Center for Free-Electron Laser Science / DESY, Notkestra\ss e 85, D-22607 Hamburg, Germany}
\affiliation{Inst. of Optics and Atomic Physics, TU Berlin, Stra\ss e des 17. Juni 135, D-10623 Berlin, Germany}

\date{\today}

\begin{abstract}
The electronic properties of co-evaporated mixtures (blends) of manganese phthalocyanine and the fullerene C$_{60}$ (MnPc\,:\,C$_{60}$) have been studied as a function of the concentration of the two constituents using two supplementary electron spectroscopic methods, photoemission spectroscopy (PES) as well as electron energy-loss spectroscopy (EELS) in transmission. Our PES measurements provide a detailed picture of the electronic structure measured with different excitation energies as well as different mixing ratios between MnPc and C$_{60}$. Besides a relative energy shift, the occupied electronic states of the two materials remain essentially unchanged. The observed energy level alignment is different compared to that of the related CuPc\,:\,C$_{60}$ bulk heterojunction. Moreover, the results from our EELS investigations show that despite of the rather small interface interaction the MnPc related electronic excitation spectrum changes significantly by admixing C$_{60}$ to MnPc thin films.
\end{abstract}

\maketitle

\section{Introduction}

Molecular thin films consisting of transition metal phthalocyaninces (Pc's) have achieved significant attention due to the fact that they play a major role in a wide variety of emerging fields of technological applications. In particular, because of their attractive semiconducting and optoelectronic properties Pc's are already being applied in various directions like, organic light-emitting diodes (OLEDs) \cite{Slyke1996,Walzer2007,Friend1999,Meerheim2009,Weichsel2012}, organic field-effect transistors (OFETs) \cite{Bao1996,Ling2006}, organic spintronic devices \cite{Dediu2002,Naber2007}, and organic photovoltaic cells (OPVs) \cite{Armstrong2009,Peumans2001,Rand2007}.

In particular, organic solar cells  have a large potential as flexible low cost material systems for photovoltaic power generation. An additional advantage of OPV systems is that they can easily be prepared by deposition from solutions, or by (low-cost) printing techniques, rather than by more demanding vacuum deposition methods. However, one major drawback is the currently quite low conversion efficiency of organic based PVs. This is mainly due to the radiative recombination of the electron-hole pair produced in the photoreceptive material by the absorption of light. If one succeeds to efficiently quench the recombination process, the yield of photogenerated charges may be significantly improved.

\begin{figure}[b]
\includegraphics[width=0.8\linewidth]{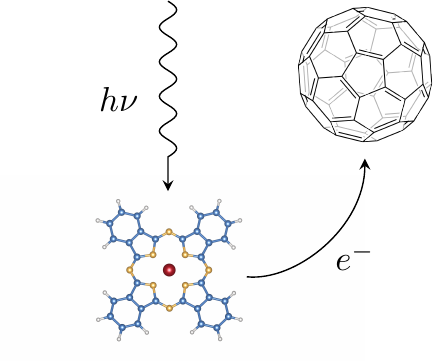}
\caption{Principle scheme of the charge transfer process in an organic heterojunction between a phthalocyanine (acting as electron-donor) and C$_{60}$ (acting as acceptor). On the left site the schematic illustration of the chemical structure of a transition metal phthalocyanine molecule is shown, whereby the red ball represent the metal center atom, the yellow sphere stands for the 8 nitrogen atoms, and the blue and grey bullets  express the carbon and hydrogen atoms, respectively.}
\label{f1}
\end{figure}

Empirically, this has been accomplished by an admixture of C$_{60}$ into the organic photoreceptor film \cite{Sariciftci1992,Schlebusch1996,Kessler1998}. Since C$_{60}$ has a very high electron affinity, it is capable of accepting an electron from the organic photoreceptor, while the hole remains in the organic system and thus the recombination process is strongly suppressed (cf. Fig.\,\ref{f1}). Studies by photoelectron spectroscopy have shown that the charge transfer process is energetically possible, both in layered systems \cite{Schlebusch1999,Morenzin1999} as well as in heterogeneous mixtures \cite{Roth2014_CuPc}. The current state of knowledge about these systems and their role in organic photovoltaic devices has been summarized recently \cite{Opitz2012,Fahlman2013}.

Manganese phthalocyanine (MnPc) is an interesting compound within the phthalocyanine family \cite{Fu2007}. For instance, MnPc is characterized by an unusual S\,=\,3/2 spin state of the Mn$^{2+}$ ion \cite{Mitra1983,Taguchi2006,Barraclough1970} and therefore it has been referred to as a typical example of a molecular magnet. Moreover, in MnPc the Mn 3\,$d$ orbitals are expected to lie close to the chemical potential \cite{Reynolds1991,Liao2005,Arrigo2007}. Therefore, MnPc plays a special role inside the group of PC's and it was demonstrated that the occupied electronic structure as well as the electronic excitation spectrum are significantly different compared to all other transition metal phthalocyanines \cite{Grobosch2011,Kraus2009,Grobosch2010,Grobosch2009}. Further on, recent calculations confirmed experimental measurements that the ionization potential of MnPc is drastically decreased compared to other Pc's (about 500\,meV smaller) \cite{Friedrich2012,Grobosch2009}.

In this contribution we present the first comprehensive investigation of the electronic properties of MnPc - C$_{60}$ bulk heterojunctions using a combination of two spectroscopic methods, photoemission spectroscopy (PES) and electron energy-loss spectroscopy (EELS). Our results provide a detailed analysis of the changes that are induced in the electronic structure depending on different mixing rations between MnPc and C$_{60}$.

\section{Experimental}

For the PES investigations we have chosen a pre-cleaned Si(100) wafer (n-doped) as substrate in order to have a well characterized reproducible substrate conditions, where the electronic properties are well established. The fullerene and phthalocyanine films were grown on such substrates by $in\,situ$ (co)evaporation of C$_{60}$ and MnPc from two spatially separated effusion cells. The growth at room temperature was monitored by a quartz crystal microbalance to ensure homogeneous and continuous films over the 5x5\,mm$^2$ Si(100) wafer. The samples were prepared in the preparation chamber (base pressure of  $<$\,2\,x\,10$^{-8}$\,mbar) connected to the analysis chamber of the iDEEAA apparatus (base pressure of  $<$\,5\,x\,10$^{-10}$\,mbar) \cite{Lupulescu2013}.

\par

Photoelectron spectra were recorded with the SCIENTA R4000 hemispherical electron spectrometer in the iDEEAA end station \cite{Lupulescu2013}. The instrument was installed at the Metrology Light Source (MLS), located in the Willy-Wien-Laboratorium of the Physikalisch-Technische Bundesanstalt (PTB) \cite{Gottwald2012}. For this work, the iDEEAA end station was installed at the Insertion Device Beamline (IDB) whereby a 30.5 period undulator of 125\,mm length (U125) serves as insertion device. The spot size on the sample is about 1.7\,mm horizontal and 0.1\,mm vertical. At a pass energy of 20\,eV the electron spectrometer was operated at an energy resolution of 15\,meV.

\par

The EELS measurements require thin samples with a thickness of only about 100\,nm. For this purposes thin films of organic compounds have been produced by thermal evaporation under high vacuum onto a single crystalline substrates (KBr) kept at room temperature in a separate vacuum chamber. During the vacuum deposition the film thickness was monitored \emph{in situ} via a quartz crystal microbalance. Subsequent to the evaporation the films are floated off in destilled water, mounted onto standard electron microscopy grids \cite{Fink1994}, incorporated into an EELS sample holder, and transferred into the EELS spectrometer \cite{Fink1989,Roth2014}.

\par

All loss function measurements were carried out using the 172\,keV spectrometer thoroughly discussed in detail in previous publications \cite{Fink1989,Roth2014}. At this high primary beam energy only singlet excitations are possible. The energy and momentum resolution was 85\,meV and 0.03\,\AA$^{-1}$ for all measurements, respectively. The EELS signal, i.\,e., the loss function Im[-1/$\epsilon(\textbf{q},\omega)$], which is proportional to the dynamic structure factor S($\textbf{q},\omega$), was determined for a small momentum transfer, $q$ = 0.1\,\AA$^{-1}$, which represents the optical limit \cite{Fink1989}.  \\

\section{Results and discussion}
\subsection{PES on MnPc:C$_{60}$ blends}

\begin{figure*}
\includegraphics[width=0.32\linewidth]{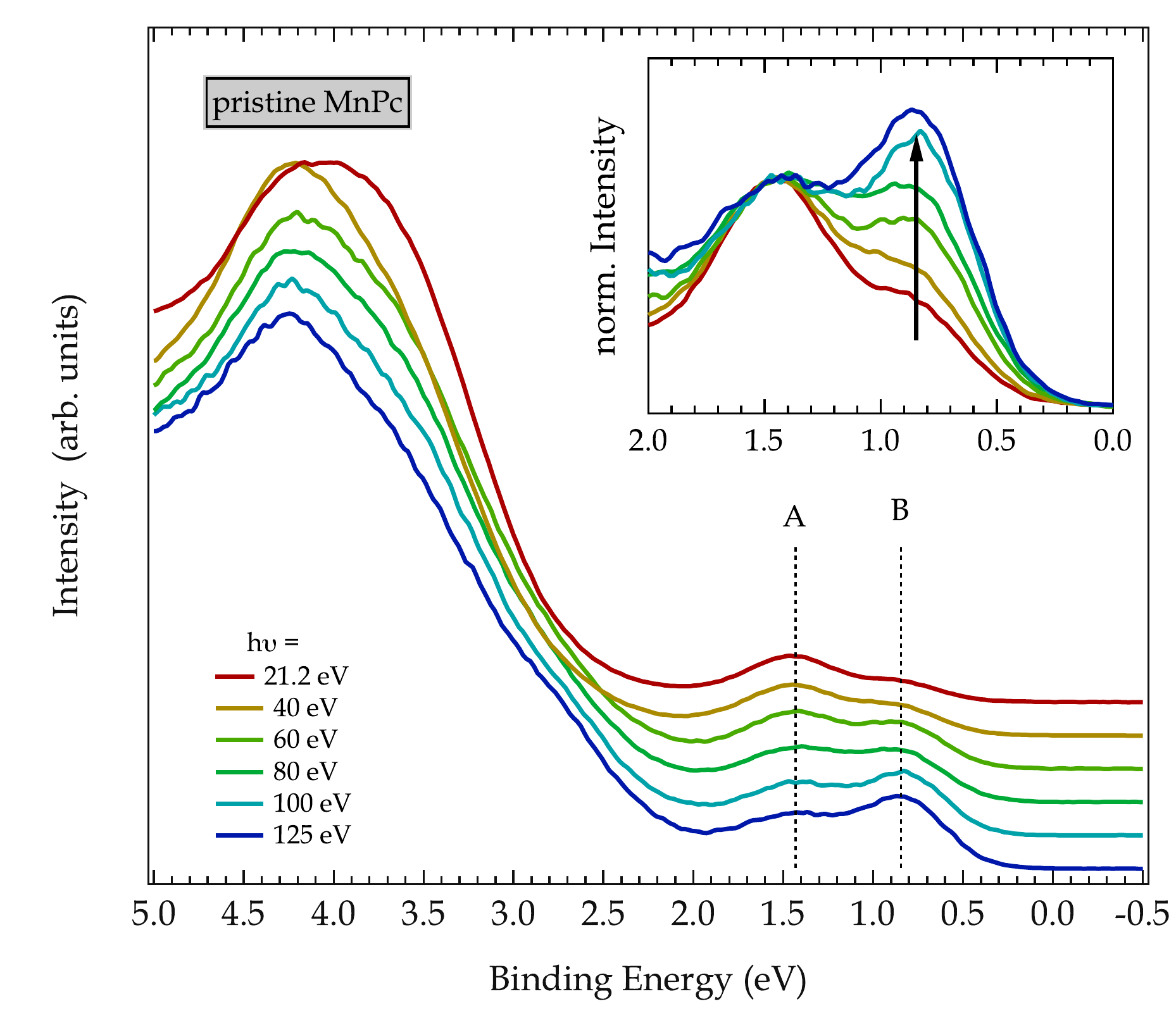}
\includegraphics[width=0.32\linewidth]{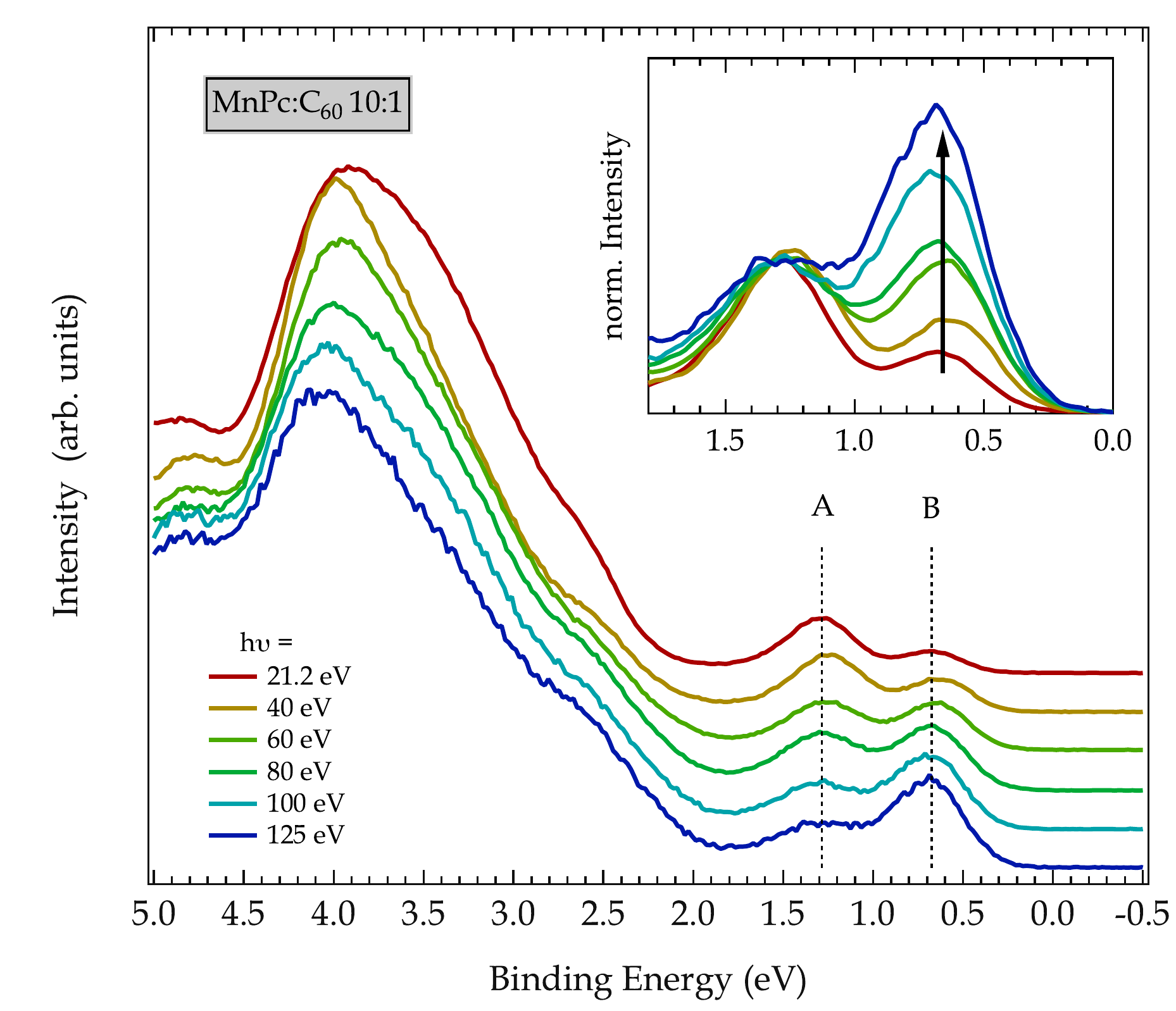}
\includegraphics[width=0.32\linewidth]{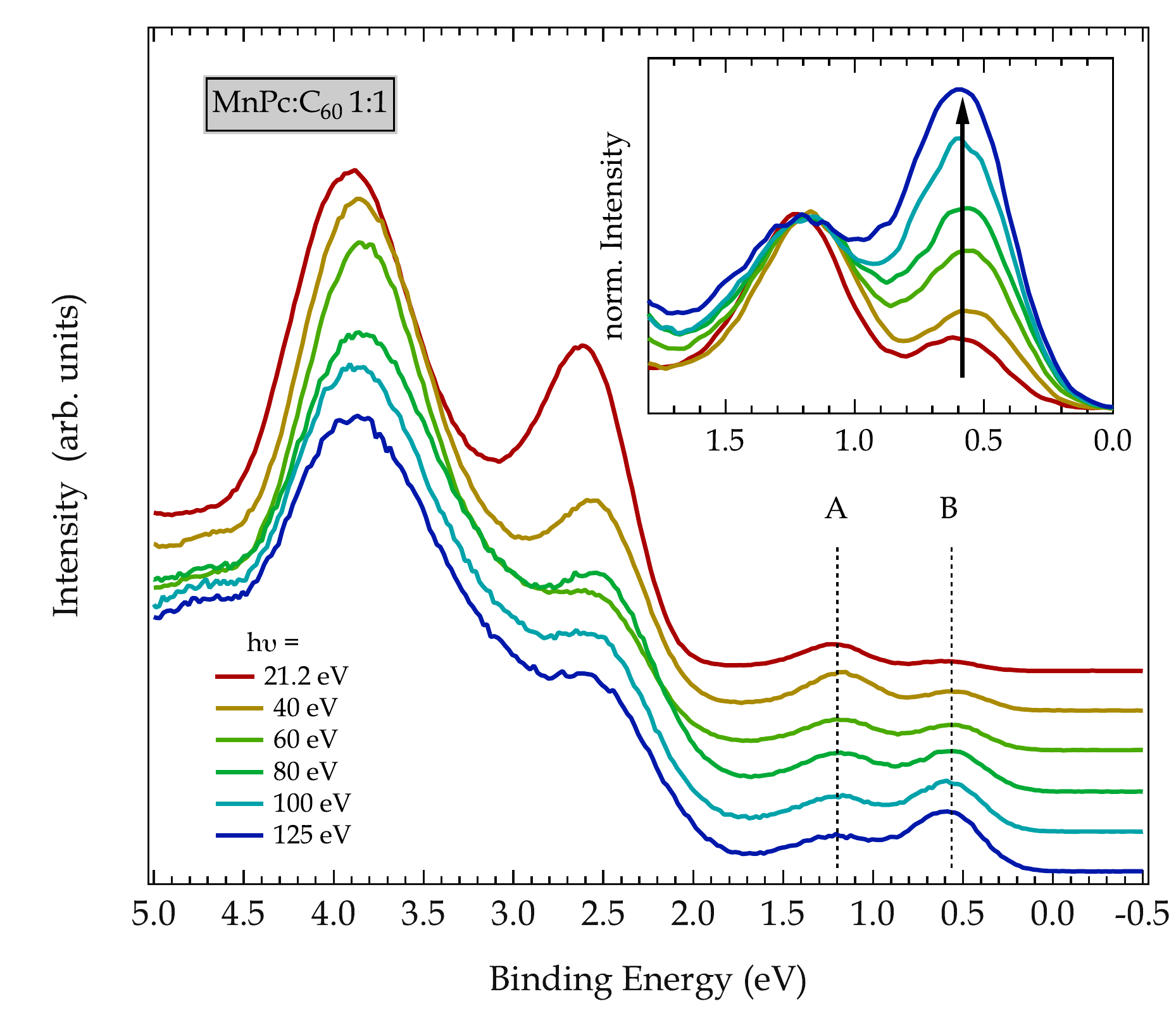}
\caption{Photoemission spectra for pristine MnPc (left panel) as well as the two composites with C$_{60}$ (10:1 - middle panel, 1:1 - right panel) measured at various photon energies ($h\nu$ increases from top to bottom). The two labels A and B highlighting the two main low-energy features of MnPc. The inset shows the intensity 	 redistribution between Peak A and B upon increasing photon energy.}
\label{f2}
\end{figure*}

\begin{figure*}[t]
\includegraphics[width=0.65\linewidth]{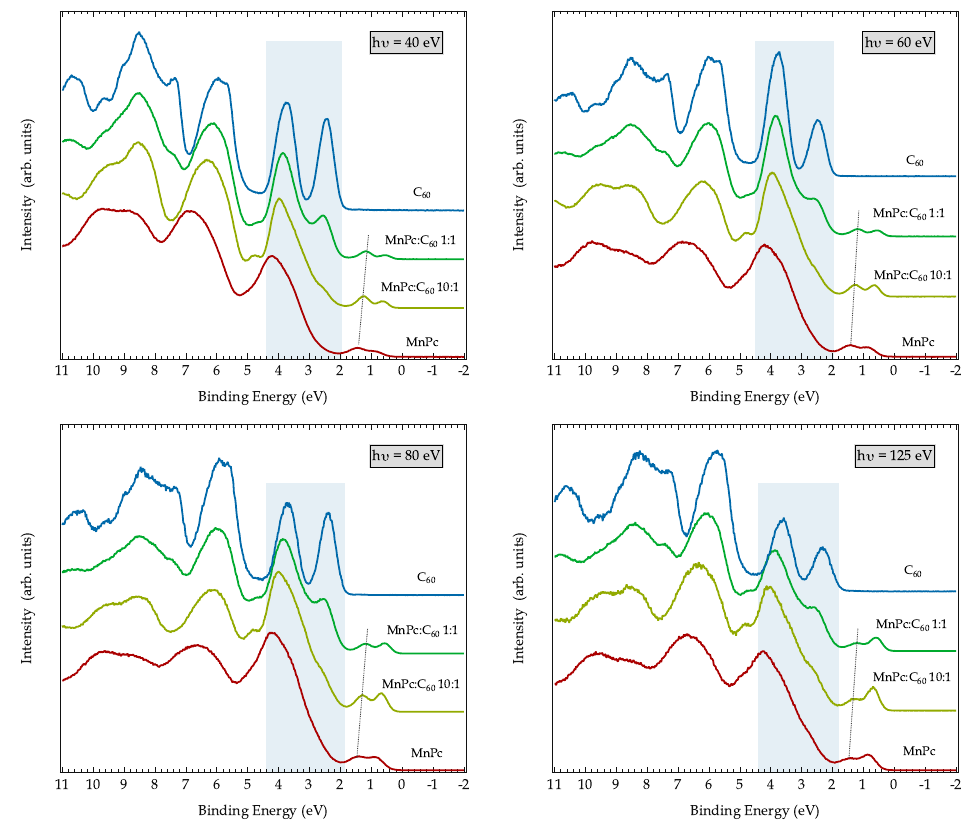}
\caption{Photoemission spectra for pristine MnPC and C$_{60}$ as well as for two mixtures of MnPc:C$_{60}$ (10:1 and 1:1) for various photon energies ($h\nu$ = 40, 60, 80, and 125\,eV). The energy scale is referenced to the Fermi level of the substrate. The blue shaded area indicates the region of the typical HOMO and HOMO-1 structure of C$_{60}$. Furthermore, the dashed lines are a guide to the eye to follow up the shifts discussed in the text.}
\label{f3}
\end{figure*}

In Figure\,\ref{f2} we show the valence band photoemission spectra for pristine MnPc, MnPc:C$_{60}$ 10:1, and  MnPc:C$_{60}$ 1:1 measured at various photon energies ($h\nu$ = 21.2, 40, 60, 80, 100, and 125\,eV). The spectra reveal the characteristic two peak low-energy structure of MnPc labeled with A and B (in the pristine compound at around 0.85\,eV and 1.45\,eV, respectively). This is in agreement with previous PES measurements on MnPc thin films \cite{Grobosch2009,Grobosch2011}. Peak A can be ascribed to emission from the $a_{1u}$ ligand state, the highest occupied molecular orbital (HOMO). The second feature closer to the chemical potential (peak B) can be only observed in MnPc and is associated to an electronic state that has significant metal 3\,$d$ character. Moreover, as shown in the insets of Fig.\,\ref{f2} the intensity of the first ionization state changes significant by increasing the excitation energy (we have normalized the spectra to the intensities of the ligand HOMO level, since this level has $a_{1u}$ symmetry and does not hybridize with any of the metal 3\,$d$ states). It grows in intensity with increasing photon energy. Interestingly, this observation is the same even for the blends of MnPc:C$_{60}$, whereby the increment of intensity seems to be more pronounced if one increase the C$_{60}$ amount (cf. Fig.\,\ref{f2} middle and right panel).

This is completely different to what we observe in CuPc (and also CuPc:C$_{60}$ mixtures), where only one low energy feature was observed without any intensity variation as a function of increasing excitation energies \cite{Roth2014_CuPc}. Even in case of copper as metal center atom that electronic state with metal 3\,$d$ character is present, but due to the different effective nuclear charge (compared to MnPc) it is situated at higher binding energies and therefore not visible in the valance band spectra. This agrees to some theoretical predictions where the metal 3\,$d$ levels move upwards by going down the transition metals series \cite{Liao2001}.

According to previous publications we attribute peak B to a Mn\,3\,$d$ state with $e_g$ symmetry, since this is the only metal 3\,$d$ state that can hybridize with the ligand $\pi^*$ states. This hybridization is necessary to explain the observed photoemission intensity at low photon energies, because of a much larger probability for emission from a carbon or nitrogen 2\,$p$ orbital than from a transition metal 3\,$d$ state at this energies \cite{Trinity2006,Gargiani2010}.  Moreover, recent theoretical calculations predict a highest molecular orbital consisting of a ligand $e_g$ state with significant metal 3\,$d$ contribution \cite{Grobosch2011,Calzolari2007}.

\par

Fig.\,\ref{f3} shows photoemission spectra of nominally 7-10\,nm thick films taken at four different photon energies ($h\nu$ = 40, 60, 80, and 125\,eV). For each photon energy, the spectra are shown for the pure MnPc, a 10:1 MnPc:C$_{60}$ film, a 1:1 MnPc:C$_{60}$ film, and a pure C$_{60}$ film. The energy scale is referenced to the Fermi level of a thin gold film, which was deposited on the Si substrate.

The changes by admixing C$_{60}$ to MnPc are twofold. First of all, the increase of the C$_{60}$ content in the thin films manifests itself by a rise of spectral weight (cf. bluish shaded area in Fig.\,\ref{f3}) in the region where pristine C$_{60}$ shows up the characteristic two peak HOMO and HOMO-1 structure \cite{Weaver1992,Golden1995}. Especially, this becomes clear by focusing on the first valence band excitation (HOMO) of C$_{60}$ around 2.5\,eV, which gets conspicuous by going from a mixture of 10:1 over 1:1 up to pure C$_{60}$.

\par

Besides this more or less expected result, the second modification is the change of the level alignment from the pure films to the mixtures/blends. This is clearly visible when observing the double peak structure of MnPc, which shifts to lower binding energies by increasing the C$_{60}$ amount. The position of the C$_{60}$ derived emission is not as easily derived especially for the low concentration (10:1) blends. To shine more light on this, we extract the shift of both parent compounds by first fitting the MnPc HOMO structure with the help of two Voigt peaks and second using the resulting shift of the MnPc and fit the spectra of the blends by using the spectra of the pure compounds allowing for an energy shift and intensity variation (whereby the energy shift of MnPc is fixed to the value of the previous fit procedure).

\par

The MnPc shifts in the 1:1 mixture are slightly larger (-240\,meV) than in the 10:1 blend (-160\,meV), while the shifts in the C$_{60}$ component are slightly smaller for the 1:1 blend (+70\,meV) compared with 165\,meV for the 10:1 blend. From this it is remarkable that the relative alignment of the two constituents of this bulk heterojunction
remains identical at about 310\,-\,325\,meV shift measured relative to the electronic states of the separate systems. Interestingly, this shift is of about 35 - 40\,\% less compared to mixed films including CuPc instead of MnPc with same ratios, where the resulting shift was about 500\,meV \cite{Roth2014_CuPc}. In general, energy shifts or the formation of dipoles are thoroughly studied phenomena at interfaces of organic semiconducting materials to metals but also at organic heterojunctions \cite{Koch2007}, which are of particular importance for applications. Recent models describe these dipoles via the charge flow across the interfaces during the formation process, which is dependent on the density of states of the organic semiconductors \cite{Vazquez2007,Flores2009,Oehzelt2014}. Moreover, these models suggest that the dipoles found at organic heterojunctions can be modeled knowing the dipoles of either organic semiconductor to inert metals (e.\,g., Au). Moreover, the interface dipoles (energy shifts) observed for a bulk heterojunction of CuPc:C$_{60}$ and a well ordered individual interface agree well \cite{Molodtsova2006,Roth2014_CuPc}. The interface dipoles at CuPc/Au and MnPc/Au interfaces can be found in the literature, and it turns out that they are different by about 300\,meV \cite{Grobosch2009}. This can be rationalized in terms of different Fermi energy (or charge neutrality level, see \cite{Vazquez2007}) positions in CuPc and MnPc 
following the IDIS (induced density of interface states) model\cite{Vazquez2007}. This difference now is also reflected at the heterojunctions to C$_{60}$ as discussed here, and leads to the observed variation in the energy shifts of the photoemission signals as presented above, i.\,e., the IDIS model is appropriate to describe the observed energy shifts.

\par

Further on, as shown in our previous paper, we assume that even in case of MnPc, instead of CuPc, the morphology changes by admixing C$_{60}$ into the system, whereas the electronic structure will not effected by this morphology variances \cite{Roth2014_CuPc}.

\subsection{EELS on MnPc:C$_{60}$ blends}

\begin{figure}[b]
\includegraphics[width=0.9\linewidth]{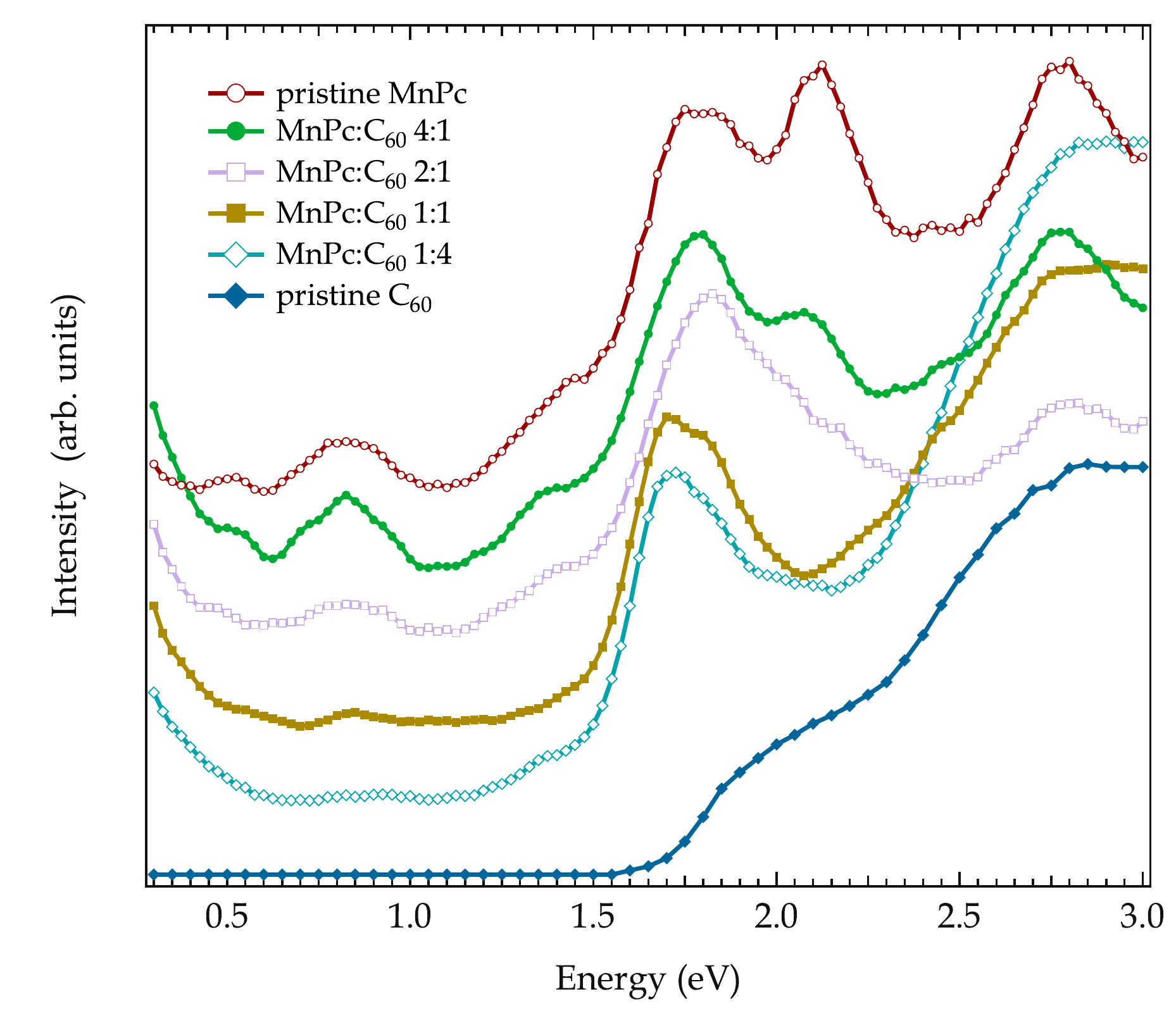}
\caption{Loss function of pristine MnPc (red, open circles) as well as C$_{60}$ (blue, full diamonds) and various heterojunctions with mixing ratios of MnPc:C$_{60}$ of 4:1 (green, full circles), 2:1 (purple, open squares), 1:1 (dark yellow, full squares), and 1:4 (turquoise, open diamonds). The upturn toward 0\,eV is due to the quasi-elastic line of the direct beam.}
\label{f4}
\end{figure}

In Fig.\,\ref{f4} we summarize the evolution of the electronic excitation spectra of MnPc:C$_{60}$ bulk heterojunctions below 3 eV for different mixing ratios. As a consequence of an optically forbidden transition between the highest occupied molecular orbital and the lowest unoccupied molecular orbital, the excitation spectrum of C$_{60}$ starts weakly around 1.8\,eV followed by shoulder like structures at about 2 eV and 2.8 eV \cite{Sohmen1992,Hartmann1995,Knupfer1999}. It completely overlaps with the excitation spectrum of MnPc, which renders the analysis of individual features rather inaccurate.

The electronic excitation spectrum of pure MnPc is rather complex with several maxima and shoulders below 1.6\,eV and, in this respect, also differs from those of other metal phthalocyanines \cite{Fielding1964,Kraus2009}. Spectral features at about 2.1\,eV, 1.75\,eV, 1.4\,eV, 0.8\,eV, and 0.5\,eV can be observed. We note that a full microscopic understanding of this spectrum has not been achieved yet. 
Partly, the fact that Mn\,3$d$ states are close to the Fermi level in MnPc can explain the difference to other phthalocyanines as these states give rise to excitations below the well-known Q band (around 1.8 to 2\,eV) of the phthalocynines \cite{Grobosch2010}.

\par

Going to the MnPc:C$_{60}$ mixed films, the MnPc derived excitation spectrum changes quite drastically. With increasing C$_{60}$ admixture the excitation features at 0.5\,eV and 2.1\,eV  are lost. The excitations at 0.8\,eV as well as the shoulder around 1.4\,eV decrease in intensity and become somewhat narrower. Finally, the remaining excitation for almost diluted MnPc in a C$_{60}$  matrix (mixing ratio 1:4) resembles very much that of individual MnPc molecules in solution \cite{Engelsma1962,Lever1981}.

\begin{figure}[t]
\includegraphics[width=0.9\linewidth]{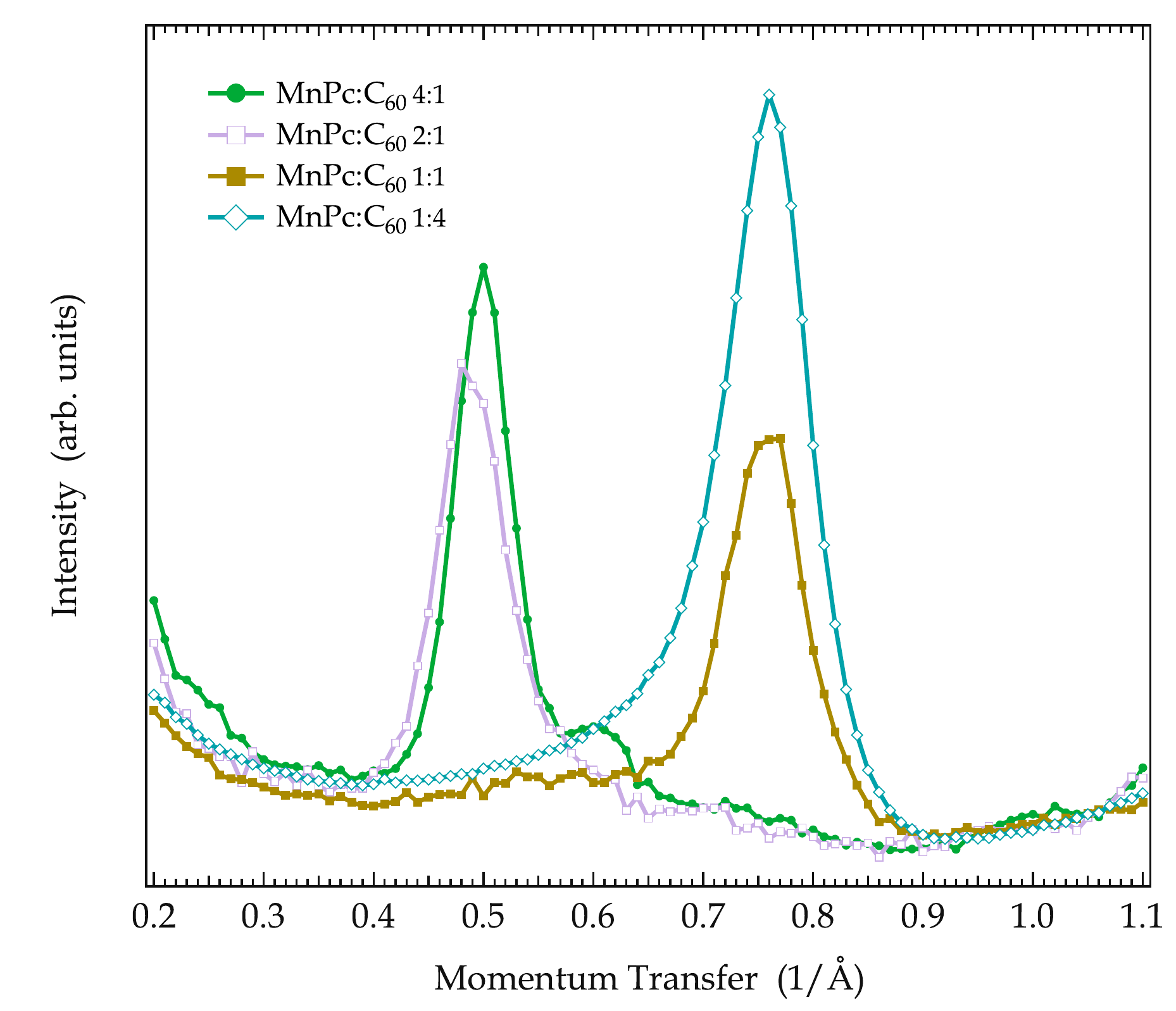}
\caption{Comparison of the electron diffraction profiles of MnPc:C$_{60}$ blends with different mixing ratios in the range between 0.2 and 1.1 \AA$^{-1}$. The annotation is the same as in Fig.\,\ref{f4}.}
\label{f5}
\end{figure}

For many phthalocyanines it is known that the electronic excitation spectrum in the bulk is rather different from that in solution, in monolayers or in the gas phase \cite{Sharp1973,Yoshida1986,Auerhammer2002,Gunaratne2004}. This difference had been attributed to the interaction of the molecules in the excited state (cf. Davydov splitting), a Jahn-Teller like distortion of the molecules, and the admixture of so-called charge-transfer excitations to the excited state wave function \cite{Yoshida1986,Knupfer2004}. This might also explain the complexity of the MnPc bulk excitation spectra where, in addition to the $\pi$-derived ligand orbitals,  Mn\,3$d$ derived states participate at low energy \cite{Grobosch2011}. Adding more and more of C$_{60}$ results in smaller and smaller MnPc aggregates/grains and less intermolecular impact on the excitation spectra. This behavior has already been observed for CuPc:C$_{60}$ mixed films \cite{Rand2005}, and it is also represented in the electron diffraction profiles of the mixed films shown in Fig.\,\ref{f5}. The diffraction peak at about 0.5\,\AA$^{-1}$ is due to the (200) Bragg reflection of the phthalocyanine \cite{Rand2005}. It disappears for mixing ratios of 1:1 and 1:4, in good agreement to what has been observed for CuPc:C$_{60}$ films \cite{Rand2005}. Simultaneously, the C$_{60}$ derived (111) Bragg reflection shows up at about 0.76\,eV, which was absent for larger MnPc:C$_{60}$ ratios. This indicates that for an admixture of more than about 30\,\% of C$_{60}$ no sizable crystalline MnPc grains are formed any more, while at lower C$_{60}$ content this is true for C$_{60}$.

\section{Conclusion}
We have studied mixtures (blends) of MnPc and C$_{60}$ as model systems for bulk heterojunction organic materials using two supplementary electron spectroscopic methods, photoemission spectroscopy (PES) as well as electron energy-loss spectroscopy (EELS) in transmission. The main objective of our research was to shine more light on the role of MnPc in the series of transition metal phthalocyanines with respect to interfacial as well as bulk properties. Our PES measurements provide an extensive overview about the change of the electronic structure in the valence band region upon admixing  C$_{60}$ to MnPc. Furthermore, an electronic interface dipole of about 300\,meV is observed, whereby the relative band alignment keeps essentially unchanged. Interestingly, this value is significantly smaller compared to CuPc:C$_{60}$ mixtures.  Additionally, the EELS measurements supply a detailed picture of the drastic change in the absorptivity of the bulk heterojunction.

\begin{acknowledgments}
The photoemission studies were performed at the Metrology Light Source (MLS) of the Physikalisch-Technische Bundesanstalt (PTB) in Berlin. We would like to thank the staff of the MLS, especially H. Kaser  for the experimental support. Moreover, we thank M. Naumann, R. H\"ubel and S. Leger for technical assistance for the measurements at the IFW Dresden. Part of this work has been supported by the Deutsche Forschungsgemeinschaft under KN393/13 and KN393/14.
\end{acknowledgments}


\begin{thebibliography}{60}%
\makeatletter
\providecommand \@ifxundefined [1]{%
 \@ifx{#1\undefined}
}%
\providecommand \@ifnum [1]{%
 \ifnum #1\expandafter \@firstoftwo
 \else \expandafter \@secondoftwo
 \fi
}%
\providecommand \@ifx [1]{%
 \ifx #1\expandafter \@firstoftwo
 \else \expandafter \@secondoftwo
 \fi
}%
\providecommand \natexlab [1]{#1}%
\providecommand \enquote  [1]{``#1''}%
\providecommand \bibnamefont  [1]{#1}%
\providecommand \bibfnamefont [1]{#1}%
\providecommand \citenamefont [1]{#1}%
\providecommand \href@noop [0]{\@secondoftwo}%
\providecommand \href [0]{\begingroup \@sanitize@url \@href}%
\providecommand \@href[1]{\@@startlink{#1}\@@href}%
\providecommand \@@href[1]{\endgroup#1\@@endlink}%
\providecommand \@sanitize@url [0]{\catcode `\\12\catcode `\$12\catcode
  `\&12\catcode `\#12\catcode `\^12\catcode `\_12\catcode `\%12\relax}%
\providecommand \@@startlink[1]{}%
\providecommand \@@endlink[0]{}%
\providecommand \url  [0]{\begingroup\@sanitize@url \@url }%
\providecommand \@url [1]{\endgroup\@href {#1}{\urlprefix }}%
\providecommand \urlprefix  [0]{URL }%
\providecommand \Eprint [0]{\href }%
\providecommand \doibase [0]{http://dx.doi.org/}%
\providecommand \selectlanguage [0]{\@gobble}%
\providecommand \bibinfo  [0]{\@secondoftwo}%
\providecommand \bibfield  [0]{\@secondoftwo}%
\providecommand \translation [1]{[#1]}%
\providecommand \BibitemOpen [0]{}%
\providecommand \bibitemStop [0]{}%
\providecommand \bibitemNoStop [0]{.\EOS\space}%
\providecommand \EOS [0]{\spacefactor3000\relax}%
\providecommand \BibitemShut  [1]{\csname bibitem#1\endcsname}%
\let\auto@bib@innerbib\@empty
\bibitem [{\citenamefont {Van~Slyke}\ \emph {et~al.}(1996)\citenamefont
  {Van~Slyke}, \citenamefont {Chen},\ and\ \citenamefont {Tang}}]{Slyke1996}%
  \BibitemOpen
  \bibfield  {author} {\bibinfo {author} {\bibfnamefont {S.~A.}\ \bibnamefont
  {Van~Slyke}}, \bibinfo {author} {\bibfnamefont {C.~H.}\ \bibnamefont {Chen}},
  \ and\ \bibinfo {author} {\bibfnamefont {C.~W.}\ \bibnamefont {Tang}},\
  }\href@noop {} {\bibfield  {journal} {\bibinfo  {journal} {Appl. Phys.
  Lett.}\ }\textbf {\bibinfo {volume} {69}} (\bibinfo {year}
  {1996})}\BibitemShut {NoStop}%
\bibitem [{\citenamefont {Walzer}\ \emph {et~al.}(2007)\citenamefont {Walzer},
  \citenamefont {Maennig}, \citenamefont {Pfeiffer},\ and\ \citenamefont
  {Leo}}]{Walzer2007}%
  \BibitemOpen
  \bibfield  {author} {\bibinfo {author} {\bibfnamefont {K.}~\bibnamefont
  {Walzer}}, \bibinfo {author} {\bibfnamefont {B.}~\bibnamefont {Maennig}},
  \bibinfo {author} {\bibfnamefont {M.}~\bibnamefont {Pfeiffer}}, \ and\
  \bibinfo {author} {\bibfnamefont {K.}~\bibnamefont {Leo}},\ }\href {\doibase
  10.1021/cr050156n} {\bibfield  {journal} {\bibinfo  {journal} {Chem. Rev.}\
  }\textbf {\bibinfo {volume} {107}},\ \bibinfo {pages} {1233} (\bibinfo {year}
  {2007})}\BibitemShut {NoStop}%
\bibitem [{\citenamefont {Friend}\ \emph {et~al.}(1999)\citenamefont {Friend},
  \citenamefont {Gymer}, \citenamefont {Holmes}, \citenamefont {Burroughes},
  \citenamefont {Marks}, \citenamefont {Taliani}, \citenamefont {Bradley},
  \citenamefont {Santos}, \citenamefont {Bredas}, \citenamefont {Logdlund},\
  and\ \citenamefont {Salaneck}}]{Friend1999}%
  \BibitemOpen
  \bibfield  {author} {\bibinfo {author} {\bibfnamefont {R.~H.}\ \bibnamefont
  {Friend}}, \bibinfo {author} {\bibfnamefont {R.~W.}\ \bibnamefont {Gymer}},
  \bibinfo {author} {\bibfnamefont {A.~B.}\ \bibnamefont {Holmes}}, \bibinfo
  {author} {\bibfnamefont {J.~H.}\ \bibnamefont {Burroughes}}, \bibinfo
  {author} {\bibfnamefont {R.~N.}\ \bibnamefont {Marks}}, \bibinfo {author}
  {\bibfnamefont {C.}~\bibnamefont {Taliani}}, \bibinfo {author} {\bibfnamefont
  {D.~D.~C.}\ \bibnamefont {Bradley}}, \bibinfo {author} {\bibfnamefont
  {D.~A.~D.}\ \bibnamefont {Santos}}, \bibinfo {author} {\bibfnamefont {J.~L.}\
  \bibnamefont {Bredas}}, \bibinfo {author} {\bibfnamefont {M.}~\bibnamefont
  {Logdlund}}, \ and\ \bibinfo {author} {\bibfnamefont {W.~R.}\ \bibnamefont
  {Salaneck}},\ }\href {http://dx.doi.org/10.1038/16393} {\bibfield  {journal}
  {\bibinfo  {journal} {Nature}\ }\textbf {\bibinfo {volume} {397}},\ \bibinfo
  {pages} {121} (\bibinfo {year} {1999})}\BibitemShut {NoStop}%
\bibitem [{\citenamefont {Meerheim}\ \emph {et~al.}(2009)\citenamefont
  {Meerheim}, \citenamefont {L\"ussem},\ and\ \citenamefont
  {Leo}}]{Meerheim2009}%
  \BibitemOpen
  \bibfield  {author} {\bibinfo {author} {\bibfnamefont {R.}~\bibnamefont
  {Meerheim}}, \bibinfo {author} {\bibfnamefont {B.}~\bibnamefont {L\"ussem}},
  \ and\ \bibinfo {author} {\bibfnamefont {K.}~\bibnamefont {Leo}},\ }\href
  {\doibase 10.1109/JPROC.2009.2022418} {\bibfield  {journal} {\bibinfo
  {journal} {Proc. IEEE}\ }\textbf {\bibinfo {volume} {97}},\ \bibinfo {pages}
  {1606} (\bibinfo {year} {2009})}\BibitemShut {NoStop}%
\bibitem [{\citenamefont {Weichsel}\ \emph {et~al.}(2012)\citenamefont
  {Weichsel}, \citenamefont {Reineke}, \citenamefont {Furno}, \citenamefont
  {L\"ussem},\ and\ \citenamefont {Leo}}]{Weichsel2012}%
  \BibitemOpen
  \bibfield  {author} {\bibinfo {author} {\bibfnamefont {C.}~\bibnamefont
  {Weichsel}}, \bibinfo {author} {\bibfnamefont {S.}~\bibnamefont {Reineke}},
  \bibinfo {author} {\bibfnamefont {M.}~\bibnamefont {Furno}}, \bibinfo
  {author} {\bibfnamefont {B.}~\bibnamefont {L\"ussem}}, \ and\ \bibinfo
  {author} {\bibfnamefont {K.}~\bibnamefont {Leo}},\ }\href {\doibase
  http://dx.doi.org/10.1063/1.3679549} {\bibfield  {journal} {\bibinfo
  {journal} {J. Appl. Phys.}\ }\textbf {\bibinfo {volume} {111}},\ \bibinfo
  {eid} {033102} (\bibinfo {year} {2012})}\BibitemShut {NoStop}%
\bibitem [{\citenamefont {Bao}\ \emph {et~al.}(1996)\citenamefont {Bao},
  \citenamefont {Lovinger},\ and\ \citenamefont {Dodabalapur}}]{Bao1996}%
  \BibitemOpen
  \bibfield  {author} {\bibinfo {author} {\bibfnamefont {Z.}~\bibnamefont
  {Bao}}, \bibinfo {author} {\bibfnamefont {A.~J.}\ \bibnamefont {Lovinger}}, \
  and\ \bibinfo {author} {\bibfnamefont {A.}~\bibnamefont {Dodabalapur}},\
  }\href@noop {} {\bibfield  {journal} {\bibinfo  {journal} {Appl. Phys.
  Lett.}\ }\textbf {\bibinfo {volume} {69}} (\bibinfo {year}
  {1996})}\BibitemShut {NoStop}%
\bibitem [{\citenamefont {Ling}\ and\ \citenamefont {Bao}(2006)}]{Ling2006}%
  \BibitemOpen
  \bibfield  {author} {\bibinfo {author} {\bibfnamefont {M.-M.}\ \bibnamefont
  {Ling}}\ and\ \bibinfo {author} {\bibfnamefont {Z.}~\bibnamefont {Bao}},\
  }\href {\doibase http://dx.doi.org/10.1016/j.orgel.2006.09.003} {\bibfield
  {journal} {\bibinfo  {journal} {Org. Electron.}\ }\textbf {\bibinfo {volume}
  {7}},\ \bibinfo {pages} {568 } (\bibinfo {year} {2006})}\BibitemShut
  {NoStop}%
\bibitem [{\citenamefont {Dediu}\ \emph {et~al.}(2002)\citenamefont {Dediu},
  \citenamefont {Murgia}, \citenamefont {Matacotta}, \citenamefont {Taliani},\
  and\ \citenamefont {Barbanera}}]{Dediu2002}%
  \BibitemOpen
  \bibfield  {author} {\bibinfo {author} {\bibfnamefont {V.}~\bibnamefont
  {Dediu}}, \bibinfo {author} {\bibfnamefont {M.}~\bibnamefont {Murgia}},
  \bibinfo {author} {\bibfnamefont {F.}~\bibnamefont {Matacotta}}, \bibinfo
  {author} {\bibfnamefont {C.}~\bibnamefont {Taliani}}, \ and\ \bibinfo
  {author} {\bibfnamefont {S.}~\bibnamefont {Barbanera}},\ }\href {\doibase
  http://dx.doi.org/10.1016/S0038-1098(02)00090-X} {\bibfield  {journal}
  {\bibinfo  {journal} {Solid State Commun.}\ }\textbf {\bibinfo {volume}
  {122}},\ \bibinfo {pages} {181 } (\bibinfo {year} {2002})}\BibitemShut
  {NoStop}%
\bibitem [{\citenamefont {Naber}\ \emph {et~al.}(2007)\citenamefont {Naber},
  \citenamefont {Faez},\ and\ \citenamefont {van~der Wiel}}]{Naber2007}%
  \BibitemOpen
  \bibfield  {author} {\bibinfo {author} {\bibfnamefont {W.~J.~M.}\
  \bibnamefont {Naber}}, \bibinfo {author} {\bibfnamefont {S.}~\bibnamefont
  {Faez}}, \ and\ \bibinfo {author} {\bibfnamefont {W.~G.}\ \bibnamefont
  {van~der Wiel}},\ }\href {http://stacks.iop.org/0022-3727/40/i=12/a=R01}
  {\bibfield  {journal} {\bibinfo  {journal} {J. Phys. D: Appl. Phys.}\
  }\textbf {\bibinfo {volume} {40}},\ \bibinfo {pages} {R205} (\bibinfo {year}
  {2007})}\BibitemShut {NoStop}%
\bibitem [{\citenamefont {Armstrong}\ \emph {et~al.}(2009)\citenamefont
  {Armstrong}, \citenamefont {Wang}, \citenamefont {Alloway}, \citenamefont
  {Placencia}, \citenamefont {Ratcliff},\ and\ \citenamefont
  {Brumbach}}]{Armstrong2009}%
  \BibitemOpen
  \bibfield  {author} {\bibinfo {author} {\bibfnamefont {N.~R.}\ \bibnamefont
  {Armstrong}}, \bibinfo {author} {\bibfnamefont {W.}~\bibnamefont {Wang}},
  \bibinfo {author} {\bibfnamefont {D.~M.}\ \bibnamefont {Alloway}}, \bibinfo
  {author} {\bibfnamefont {D.}~\bibnamefont {Placencia}}, \bibinfo {author}
  {\bibfnamefont {E.}~\bibnamefont {Ratcliff}}, \ and\ \bibinfo {author}
  {\bibfnamefont {M.}~\bibnamefont {Brumbach}},\ }\href {\doibase
  10.1002/marc.200900075} {\bibfield  {journal} {\bibinfo  {journal} {Macromol.
  Rapid Commun.}\ }\textbf {\bibinfo {volume} {30}},\ \bibinfo {pages} {717}
  (\bibinfo {year} {2009})}\BibitemShut {NoStop}%
\bibitem [{\citenamefont {Peumans}\ and\ \citenamefont
  {Forrest}(2001)}]{Peumans2001}%
  \BibitemOpen
  \bibfield  {author} {\bibinfo {author} {\bibfnamefont {P.}~\bibnamefont
  {Peumans}}\ and\ \bibinfo {author} {\bibfnamefont {S.~R.}\ \bibnamefont
  {Forrest}},\ }\href@noop {} {\bibfield  {journal} {\bibinfo  {journal} {Appl.
  Phys. Lett.}\ }\textbf {\bibinfo {volume} {79}} (\bibinfo {year}
  {2001})}\BibitemShut {NoStop}%
\bibitem [{\citenamefont {Rand}\ \emph {et~al.}(2007)\citenamefont {Rand},
  \citenamefont {Genoe}, \citenamefont {Heremans},\ and\ \citenamefont
  {Poortmans}}]{Rand2007}%
  \BibitemOpen
  \bibfield  {author} {\bibinfo {author} {\bibfnamefont {B.~P.}\ \bibnamefont
  {Rand}}, \bibinfo {author} {\bibfnamefont {J.}~\bibnamefont {Genoe}},
  \bibinfo {author} {\bibfnamefont {P.}~\bibnamefont {Heremans}}, \ and\
  \bibinfo {author} {\bibfnamefont {J.}~\bibnamefont {Poortmans}},\ }\href
  {\doibase 10.1002/pip.788} {\bibfield  {journal} {\bibinfo  {journal} {Prog.
  Photovolt.}\ }\textbf {\bibinfo {volume} {15}},\ \bibinfo {pages} {659}
  (\bibinfo {year} {2007})}\BibitemShut {NoStop}%
\bibitem [{\citenamefont {Sariciftci}\ \emph {et~al.}(1992)\citenamefont
  {Sariciftci}, \citenamefont {Smilowitz}, \citenamefont {Heeger},\ and\
  \citenamefont {Wudl}}]{Sariciftci1992}%
  \BibitemOpen
  \bibfield  {author} {\bibinfo {author} {\bibfnamefont {N.~S.}\ \bibnamefont
  {Sariciftci}}, \bibinfo {author} {\bibfnamefont {L.}~\bibnamefont
  {Smilowitz}}, \bibinfo {author} {\bibfnamefont {A.~J.}\ \bibnamefont
  {Heeger}}, \ and\ \bibinfo {author} {\bibfnamefont {F.}~\bibnamefont
  {Wudl}},\ }\href {\doibase 10.1126/science.258.5087.1474} {\bibfield
  {journal} {\bibinfo  {journal} {Science}\ }\textbf {\bibinfo {volume}
  {258}},\ \bibinfo {pages} {1474} (\bibinfo {year} {1992})}\BibitemShut
  {NoStop}%
\bibitem [{\citenamefont {Schlebusch}\ \emph {et~al.}(1996)\citenamefont
  {Schlebusch}, \citenamefont {Kessler}, \citenamefont {Cramm},\ and\
  \citenamefont {Eberhardt}}]{Schlebusch1996}%
  \BibitemOpen
  \bibfield  {author} {\bibinfo {author} {\bibfnamefont {C.}~\bibnamefont
  {Schlebusch}}, \bibinfo {author} {\bibfnamefont {B.}~\bibnamefont {Kessler}},
  \bibinfo {author} {\bibfnamefont {S.}~\bibnamefont {Cramm}}, \ and\ \bibinfo
  {author} {\bibfnamefont {W.}~\bibnamefont {Eberhardt}},\ }\href {\doibase
  http://dx.doi.org/10.1016/0379-6779(96)80077-4} {\bibfield  {journal}
  {\bibinfo  {journal} {Synth. Met.}\ }\textbf {\bibinfo {volume} {77}},\
  \bibinfo {pages} {151 } (\bibinfo {year} {1996})}\BibitemShut {NoStop}%
\bibitem [{\citenamefont {Kessler}\ \emph {et~al.}(1998)\citenamefont
  {Kessler}, \citenamefont {Schlebusch}, \citenamefont {Morenzin},\ and\
  \citenamefont {Eberhardt}}]{Kessler1998}%
  \BibitemOpen
  \bibfield  {author} {\bibinfo {author} {\bibfnamefont {B.}~\bibnamefont
  {Kessler}}, \bibinfo {author} {\bibfnamefont {C.}~\bibnamefont {Schlebusch}},
  \bibinfo {author} {\bibfnamefont {J.}~\bibnamefont {Morenzin}}, \ and\
  \bibinfo {author} {\bibfnamefont {W.}~\bibnamefont {Eberhardt}},\ }\href@noop
  {} {\bibfield  {journal} {\bibinfo  {journal} {AIP Conf. Proc.}\ }\textbf
  {\bibinfo {volume} {442}} (\bibinfo {year} {1998})}\BibitemShut {NoStop}%
\bibitem [{\citenamefont {Schlebusch}\ \emph {et~al.}(1999)\citenamefont
  {Schlebusch}, \citenamefont {Morenzin}, \citenamefont {Kessler},\ and\
  \citenamefont {Eberhardt}}]{Schlebusch1999}%
  \BibitemOpen
  \bibfield  {author} {\bibinfo {author} {\bibfnamefont {C.}~\bibnamefont
  {Schlebusch}}, \bibinfo {author} {\bibfnamefont {J.}~\bibnamefont
  {Morenzin}}, \bibinfo {author} {\bibfnamefont {B.}~\bibnamefont {Kessler}}, \
  and\ \bibinfo {author} {\bibfnamefont {W.}~\bibnamefont {Eberhardt}},\ }\href
  {\doibase http://dx.doi.org/10.1016/S0008-6223(98)00260-7} {\bibfield
  {journal} {\bibinfo  {journal} {Carbon}\ }\textbf {\bibinfo {volume} {37}},\
  \bibinfo {pages} {717 } (\bibinfo {year} {1999})}\BibitemShut {NoStop}%
\bibitem [{\citenamefont {Morenzin}\ \emph {et~al.}(1999)\citenamefont
  {Morenzin}, \citenamefont {Schlebusch}, \citenamefont {Kessler},\ and\
  \citenamefont {Eberhardt}}]{Morenzin1999}%
  \BibitemOpen
  \bibfield  {author} {\bibinfo {author} {\bibfnamefont {J.}~\bibnamefont
  {Morenzin}}, \bibinfo {author} {\bibfnamefont {C.}~\bibnamefont
  {Schlebusch}}, \bibinfo {author} {\bibfnamefont {B.}~\bibnamefont {Kessler}},
  \ and\ \bibinfo {author} {\bibfnamefont {W.}~\bibnamefont {Eberhardt}},\
  }\href {\doibase 10.1039/A808599D} {\bibfield  {journal} {\bibinfo  {journal}
  {Phys. Chem. Chem. Phys.}\ }\textbf {\bibinfo {volume} {1}},\ \bibinfo
  {pages} {1765} (\bibinfo {year} {1999})}\BibitemShut {NoStop}%
\bibitem [{\citenamefont {Roth}\ \emph
  {et~al.}(2014{\natexlab{a}})\citenamefont {Roth}, \citenamefont {Lupulescu},
  \citenamefont {Arion}, \citenamefont {Darlatt}, \citenamefont {Gottwald},\
  and\ \citenamefont {Eberhardt}}]{Roth2014_CuPc}%
  \BibitemOpen
  \bibfield  {author} {\bibinfo {author} {\bibfnamefont {F.}~\bibnamefont
  {Roth}}, \bibinfo {author} {\bibfnamefont {C.}~\bibnamefont {Lupulescu}},
  \bibinfo {author} {\bibfnamefont {T.}~\bibnamefont {Arion}}, \bibinfo
  {author} {\bibfnamefont {E.}~\bibnamefont {Darlatt}}, \bibinfo {author}
  {\bibfnamefont {A.}~\bibnamefont {Gottwald}}, \ and\ \bibinfo {author}
  {\bibfnamefont {W.}~\bibnamefont {Eberhardt}},\ }\href {\doibase
  http://dx.doi.org/10.1063/1.4861886} {\bibfield  {journal} {\bibinfo
  {journal} {J. Appl. Phys.}\ }\textbf {\bibinfo {volume} {115}},\ \bibinfo
  {eid} {033705} (\bibinfo {year} {2014}{\natexlab{a}})}\BibitemShut {NoStop}%
\bibitem [{\citenamefont {Opitz}\ \emph {et~al.}(2013)\citenamefont {Opitz},
  \citenamefont {Frisch}, \citenamefont {Schlesinger}, \citenamefont {Wilke},\
  and\ \citenamefont {Koch}}]{Opitz2012}%
  \BibitemOpen
  \bibfield  {author} {\bibinfo {author} {\bibfnamefont {A.}~\bibnamefont
  {Opitz}}, \bibinfo {author} {\bibfnamefont {J.}~\bibnamefont {Frisch}},
  \bibinfo {author} {\bibfnamefont {R.}~\bibnamefont {Schlesinger}}, \bibinfo
  {author} {\bibfnamefont {A.}~\bibnamefont {Wilke}}, \ and\ \bibinfo {author}
  {\bibfnamefont {N.}~\bibnamefont {Koch}},\ }\href
  {http://www.sciencedirect.com/science/article/pii/S0368204812001466}
  {\bibfield  {journal} {\bibinfo  {journal} {J. Electron. Spectrosc. Relat.
  Phenom.}\ }\textbf {\bibinfo {volume} {190}},\ \bibinfo {pages} {12}
  (\bibinfo {year} {2013})},\ \bibinfo {note} {and references
  therein}\BibitemShut {NoStop}%
\bibitem [{\citenamefont {Fahlman}\ \emph {et~al.}(2013)\citenamefont
  {Fahlman}, \citenamefont {Sehati}, \citenamefont {Osikowicz}, \citenamefont
  {Braun}, \citenamefont {de~Jong},\ and\ \citenamefont
  {Brocks}}]{Fahlman2013}%
  \BibitemOpen
  \bibfield  {author} {\bibinfo {author} {\bibfnamefont {M.}~\bibnamefont
  {Fahlman}}, \bibinfo {author} {\bibfnamefont {P.}~\bibnamefont {Sehati}},
  \bibinfo {author} {\bibfnamefont {W.}~\bibnamefont {Osikowicz}}, \bibinfo
  {author} {\bibfnamefont {S.}~\bibnamefont {Braun}}, \bibinfo {author}
  {\bibfnamefont {M.~P.}\ \bibnamefont {de~Jong}}, \ and\ \bibinfo {author}
  {\bibfnamefont {G.}~\bibnamefont {Brocks}},\ }\href
  {http://www.sciencedirect.com/science/article/pii/S0368204813000352}
  {\bibfield  {journal} {\bibinfo  {journal} {J. Electron. Spectrosc. Relat.
  Phenom.}\ }\textbf {\bibinfo {volume} {190}},\ \bibinfo {pages} {33}
  (\bibinfo {year} {2013})},\ \bibinfo {note} {and references
  therein}\BibitemShut {NoStop}%
\bibitem [{\citenamefont {Fu}\ \emph {et~al.}(2007)\citenamefont {Fu},
  \citenamefont {Ji}, \citenamefont {Chen}, \citenamefont {Ma}, \citenamefont
  {Wu}, \citenamefont {Wang}, \citenamefont {Duan}, \citenamefont {Qiu},
  \citenamefont {Sun}, \citenamefont {Zhang}, \citenamefont {Jia},\ and\
  \citenamefont {Xue}}]{Fu2007}%
  \BibitemOpen
  \bibfield  {author} {\bibinfo {author} {\bibfnamefont {Y.-S.}\ \bibnamefont
  {Fu}}, \bibinfo {author} {\bibfnamefont {S.-H.}\ \bibnamefont {Ji}}, \bibinfo
  {author} {\bibfnamefont {X.}~\bibnamefont {Chen}}, \bibinfo {author}
  {\bibfnamefont {X.-C.}\ \bibnamefont {Ma}}, \bibinfo {author} {\bibfnamefont
  {R.}~\bibnamefont {Wu}}, \bibinfo {author} {\bibfnamefont {C.-C.}\
  \bibnamefont {Wang}}, \bibinfo {author} {\bibfnamefont {W.-H.}\ \bibnamefont
  {Duan}}, \bibinfo {author} {\bibfnamefont {X.-H.}\ \bibnamefont {Qiu}},
  \bibinfo {author} {\bibfnamefont {B.}~\bibnamefont {Sun}}, \bibinfo {author}
  {\bibfnamefont {P.}~\bibnamefont {Zhang}}, \bibinfo {author} {\bibfnamefont
  {J.-F.}\ \bibnamefont {Jia}}, \ and\ \bibinfo {author} {\bibfnamefont
  {Q.-K.}\ \bibnamefont {Xue}},\ }\href {\doibase
  10.1103/PhysRevLett.99.256601} {\bibfield  {journal} {\bibinfo  {journal}
  {Phys. Rev. Lett.}\ }\textbf {\bibinfo {volume} {99}},\ \bibinfo {pages}
  {256601} (\bibinfo {year} {2007})}\BibitemShut {NoStop}%
\bibitem [{\citenamefont {Mitra}\ \emph {et~al.}(1983)\citenamefont {Mitra},
  \citenamefont {Gregson}, \citenamefont {Hatfield},\ and\ \citenamefont
  {Weller}}]{Mitra1983}%
  \BibitemOpen
  \bibfield  {author} {\bibinfo {author} {\bibfnamefont {S.}~\bibnamefont
  {Mitra}}, \bibinfo {author} {\bibfnamefont {A.~K.}\ \bibnamefont {Gregson}},
  \bibinfo {author} {\bibfnamefont {W.~E.}\ \bibnamefont {Hatfield}}, \ and\
  \bibinfo {author} {\bibfnamefont {R.~R.}\ \bibnamefont {Weller}},\ }\href
  {\doibase 10.1021/ic00154a007} {\bibfield  {journal} {\bibinfo  {journal}
  {Inorg. Chem.}\ }\textbf {\bibinfo {volume} {22}},\ \bibinfo {pages} {1729}
  (\bibinfo {year} {1983})}\BibitemShut {NoStop}%
\bibitem [{\citenamefont {Taguchi}\ \emph {et~al.}(2006)\citenamefont
  {Taguchi}, \citenamefont {Miyake}, \citenamefont {Margadonna}, \citenamefont
  {Kato}, \citenamefont {Prassides},\ and\ \citenamefont
  {Iwasa}}]{Taguchi2006}%
  \BibitemOpen
  \bibfield  {author} {\bibinfo {author} {\bibfnamefont {Y.}~\bibnamefont
  {Taguchi}}, \bibinfo {author} {\bibfnamefont {T.}~\bibnamefont {Miyake}},
  \bibinfo {author} {\bibfnamefont {S.}~\bibnamefont {Margadonna}}, \bibinfo
  {author} {\bibfnamefont {K.}~\bibnamefont {Kato}}, \bibinfo {author}
  {\bibfnamefont {K.}~\bibnamefont {Prassides}}, \ and\ \bibinfo {author}
  {\bibfnamefont {Y.}~\bibnamefont {Iwasa}},\ }\href {\doibase
  10.1021/ja0582657} {\bibfield  {journal} {\bibinfo  {journal} {J. Am. Chem.
  Soc.}\ }\textbf {\bibinfo {volume} {128}},\ \bibinfo {pages} {3313} (\bibinfo
  {year} {2006})}\BibitemShut {NoStop}%
\bibitem [{\citenamefont {Barraclough}\ \emph {et~al.}(1970)\citenamefont
  {Barraclough}, \citenamefont {Martin}, \citenamefont {Mitra},\ and\
  \citenamefont {Sherwood}}]{Barraclough1970}%
  \BibitemOpen
  \bibfield  {author} {\bibinfo {author} {\bibfnamefont {C.~G.}\ \bibnamefont
  {Barraclough}}, \bibinfo {author} {\bibfnamefont {R.~L.}\ \bibnamefont
  {Martin}}, \bibinfo {author} {\bibfnamefont {S.}~\bibnamefont {Mitra}}, \
  and\ \bibinfo {author} {\bibfnamefont {R.~C.}\ \bibnamefont {Sherwood}},\
  }\href@noop {} {\bibfield  {journal} {\bibinfo  {journal} {J. Chem. Phys.}\
  }\textbf {\bibinfo {volume} {53}} (\bibinfo {year} {1970})}\BibitemShut
  {NoStop}%
\bibitem [{\citenamefont {Reynolds}\ and\ \citenamefont
  {Figgis}(1991)}]{Reynolds1991}%
  \BibitemOpen
  \bibfield  {author} {\bibinfo {author} {\bibfnamefont {P.~A.}\ \bibnamefont
  {Reynolds}}\ and\ \bibinfo {author} {\bibfnamefont {B.~N.}\ \bibnamefont
  {Figgis}},\ }\href {\doibase 10.1021/ic00010a015} {\bibfield  {journal}
  {\bibinfo  {journal} {Inorg. Chem.}\ }\textbf {\bibinfo {volume} {30}},\
  \bibinfo {pages} {2294} (\bibinfo {year} {1991})}\BibitemShut {NoStop}%
\bibitem [{\citenamefont {Liao}\ \emph {et~al.}(2005)\citenamefont {Liao},
  \citenamefont {Watts},\ and\ \citenamefont {Huang}}]{Liao2005}%
  \BibitemOpen
  \bibfield  {author} {\bibinfo {author} {\bibfnamefont {M.-S.}\ \bibnamefont
  {Liao}}, \bibinfo {author} {\bibfnamefont {J.~D.}\ \bibnamefont {Watts}}, \
  and\ \bibinfo {author} {\bibfnamefont {M.-J.}\ \bibnamefont {Huang}},\ }\href
  {\doibase 10.1021/ic0401039} {\bibfield  {journal} {\bibinfo  {journal}
  {Inorg. Chem.}\ }\textbf {\bibinfo {volume} {44}},\ \bibinfo {pages} {1941}
  (\bibinfo {year} {2005})}\BibitemShut {NoStop}%
\bibitem [{\citenamefont {Calzolari}\ \emph
  {et~al.}(2007{\natexlab{a}})\citenamefont {Calzolari}, \citenamefont
  {Ferretti},\ and\ \citenamefont {Nardelli}}]{Arrigo2007}%
  \BibitemOpen
  \bibfield  {author} {\bibinfo {author} {\bibfnamefont {A.}~\bibnamefont
  {Calzolari}}, \bibinfo {author} {\bibfnamefont {A.}~\bibnamefont {Ferretti}},
  \ and\ \bibinfo {author} {\bibfnamefont {M.~B.}\ \bibnamefont {Nardelli}},\
  }\href {http://stacks.iop.org/0957-4484/18/i=42/a=424013} {\bibfield
  {journal} {\bibinfo  {journal} {Nanotechnology}\ }\textbf {\bibinfo {volume}
  {18}},\ \bibinfo {pages} {424013} (\bibinfo {year}
  {2007}{\natexlab{a}})}\BibitemShut {NoStop}%
\bibitem [{\citenamefont {Grobosch}\ \emph {et~al.}(2011)\citenamefont
  {Grobosch}, \citenamefont {Mahns}, \citenamefont {Loose}, \citenamefont
  {Friedrich}, \citenamefont {Schmidt}, \citenamefont {Kortus},\ and\
  \citenamefont {Knupfer}}]{Grobosch2011}%
  \BibitemOpen
  \bibfield  {author} {\bibinfo {author} {\bibfnamefont {M.}~\bibnamefont
  {Grobosch}}, \bibinfo {author} {\bibfnamefont {B.}~\bibnamefont {Mahns}},
  \bibinfo {author} {\bibfnamefont {C.}~\bibnamefont {Loose}}, \bibinfo
  {author} {\bibfnamefont {R.}~\bibnamefont {Friedrich}}, \bibinfo {author}
  {\bibfnamefont {C.}~\bibnamefont {Schmidt}}, \bibinfo {author} {\bibfnamefont
  {J.}~\bibnamefont {Kortus}}, \ and\ \bibinfo {author} {\bibfnamefont
  {M.}~\bibnamefont {Knupfer}},\ }\href {\doibase
  http://dx.doi.org/10.1016/j.cplett.2011.02.039} {\bibfield  {journal}
  {\bibinfo  {journal} {Chem. Phys. Lett.}\ }\textbf {\bibinfo {volume}
  {505}},\ \bibinfo {pages} {122 } (\bibinfo {year} {2011})}\BibitemShut
  {NoStop}%
\bibitem [{\citenamefont {Kraus}\ \emph {et~al.}(2009)\citenamefont {Kraus},
  \citenamefont {Grobosch},\ and\ \citenamefont {Knupfer}}]{Kraus2009}%
  \BibitemOpen
  \bibfield  {author} {\bibinfo {author} {\bibfnamefont {R.}~\bibnamefont
  {Kraus}}, \bibinfo {author} {\bibfnamefont {M.}~\bibnamefont {Grobosch}}, \
  and\ \bibinfo {author} {\bibfnamefont {M.}~\bibnamefont {Knupfer}},\ }\href
  {\doibase http://dx.doi.org/10.1016/j.cplett.2008.12.090} {\bibfield
  {journal} {\bibinfo  {journal} {Chem. Phys. Lett.}\ }\textbf {\bibinfo
  {volume} {469}},\ \bibinfo {pages} {121 } (\bibinfo {year}
  {2009})}\BibitemShut {NoStop}%
\bibitem [{\citenamefont {Grobosch}\ \emph {et~al.}(2010)\citenamefont
  {Grobosch}, \citenamefont {Schmidt}, \citenamefont {Kraus},\ and\
  \citenamefont {Knupfer}}]{Grobosch2010}%
  \BibitemOpen
  \bibfield  {author} {\bibinfo {author} {\bibfnamefont {M.}~\bibnamefont
  {Grobosch}}, \bibinfo {author} {\bibfnamefont {C.}~\bibnamefont {Schmidt}},
  \bibinfo {author} {\bibfnamefont {R.}~\bibnamefont {Kraus}}, \ and\ \bibinfo
  {author} {\bibfnamefont {M.}~\bibnamefont {Knupfer}},\ }\href {\doibase
  http://dx.doi.org/10.1016/j.orgel.2010.06.006} {\bibfield  {journal}
  {\bibinfo  {journal} {Org. Electronics}\ }\textbf {\bibinfo {volume} {11}},\
  \bibinfo {pages} {1483 } (\bibinfo {year} {2010})}\BibitemShut {NoStop}%
\bibitem [{\citenamefont {Grobosch}\ \emph {et~al.}(2009)\citenamefont
  {Grobosch}, \citenamefont {Aristov}, \citenamefont {Molodtsova},
  \citenamefont {Schmidt}, \citenamefont {Doyle}, \citenamefont {Nannarone},\
  and\ \citenamefont {Knupfer}}]{Grobosch2009}%
  \BibitemOpen
  \bibfield  {author} {\bibinfo {author} {\bibfnamefont {M.}~\bibnamefont
  {Grobosch}}, \bibinfo {author} {\bibfnamefont {V.~Y.}\ \bibnamefont
  {Aristov}}, \bibinfo {author} {\bibfnamefont {O.~V.}\ \bibnamefont
  {Molodtsova}}, \bibinfo {author} {\bibfnamefont {C.}~\bibnamefont {Schmidt}},
  \bibinfo {author} {\bibfnamefont {B.~P.}\ \bibnamefont {Doyle}}, \bibinfo
  {author} {\bibfnamefont {S.}~\bibnamefont {Nannarone}}, \ and\ \bibinfo
  {author} {\bibfnamefont {M.}~\bibnamefont {Knupfer}},\ }\href {\doibase
  10.1021/jp901731y} {\bibfield  {journal} {\bibinfo  {journal} {J. Phys. Chem.
  C}\ }\textbf {\bibinfo {volume} {113}},\ \bibinfo {pages} {13219} (\bibinfo
  {year} {2009})}\BibitemShut {NoStop}%
\bibitem [{\citenamefont {Friedrich}\ \emph {et~al.}(2012)\citenamefont
  {Friedrich}, \citenamefont {Hahn}, \citenamefont {Kortus}, \citenamefont
  {Fronk}, \citenamefont {Haidu}, \citenamefont {Salvan}, \citenamefont {Zahn},
  \citenamefont {Schlesinger}, \citenamefont {Mehring}, \citenamefont {Roth},
  \citenamefont {Mahns},\ and\ \citenamefont {Knupfer}}]{Friedrich2012}%
  \BibitemOpen
  \bibfield  {author} {\bibinfo {author} {\bibfnamefont {R.}~\bibnamefont
  {Friedrich}}, \bibinfo {author} {\bibfnamefont {T.}~\bibnamefont {Hahn}},
  \bibinfo {author} {\bibfnamefont {J.}~\bibnamefont {Kortus}}, \bibinfo
  {author} {\bibfnamefont {M.}~\bibnamefont {Fronk}}, \bibinfo {author}
  {\bibfnamefont {F.}~\bibnamefont {Haidu}}, \bibinfo {author} {\bibfnamefont
  {G.}~\bibnamefont {Salvan}}, \bibinfo {author} {\bibfnamefont {D.~R.~T.}\
  \bibnamefont {Zahn}}, \bibinfo {author} {\bibfnamefont {M.}~\bibnamefont
  {Schlesinger}}, \bibinfo {author} {\bibfnamefont {M.}~\bibnamefont
  {Mehring}}, \bibinfo {author} {\bibfnamefont {F.}~\bibnamefont {Roth}},
  \bibinfo {author} {\bibfnamefont {B.}~\bibnamefont {Mahns}}, \ and\ \bibinfo
  {author} {\bibfnamefont {M.}~\bibnamefont {Knupfer}},\ }\href {\doibase
  http://dx.doi.org/10.1063/1.3683253} {\bibfield  {journal} {\bibinfo
  {journal} {J. Chem. Phys.}\ }\textbf {\bibinfo {volume} {136}},\ \bibinfo
  {eid} {064704} (\bibinfo {year} {2012})}\BibitemShut {NoStop}%
\bibitem [{\citenamefont {Lupulescu}\ \emph {et~al.}(2013)\citenamefont
  {Lupulescu}, \citenamefont {Arion}, \citenamefont {Hergenhahn}, \citenamefont
  {Ovsyannikov}, \citenamefont {F\"orstel}, \citenamefont {Gavrila},\ and\
  \citenamefont {Eberhardt}}]{Lupulescu2013}%
  \BibitemOpen
  \bibfield  {author} {\bibinfo {author} {\bibfnamefont {C.}~\bibnamefont
  {Lupulescu}}, \bibinfo {author} {\bibfnamefont {T.}~\bibnamefont {Arion}},
  \bibinfo {author} {\bibfnamefont {U.}~\bibnamefont {Hergenhahn}}, \bibinfo
  {author} {\bibfnamefont {R.}~\bibnamefont {Ovsyannikov}}, \bibinfo {author}
  {\bibfnamefont {M.}~\bibnamefont {F\"orstel}}, \bibinfo {author}
  {\bibfnamefont {G.}~\bibnamefont {Gavrila}}, \ and\ \bibinfo {author}
  {\bibfnamefont {W.}~\bibnamefont {Eberhardt}},\ }\href {\doibase
  http://dx.doi.org/10.1016/j.elspec.2013.09.002} {\bibfield  {journal}
  {\bibinfo  {journal} {J. Electron. Spectrosc. Relat. Phenom.}\ }\textbf
  {\bibinfo {volume} {191}},\ \bibinfo {pages} {104 } (\bibinfo {year}
  {2013})}\BibitemShut {NoStop}%
\bibitem [{\citenamefont {Gottwald}\ \emph {et~al.}(2012)\citenamefont
  {Gottwald}, \citenamefont {Klein}, \citenamefont {M\"uller}, \citenamefont
  {Richter}, \citenamefont {Scholze}, \citenamefont {Thornagel},\ and\
  \citenamefont {Ulm}}]{Gottwald2012}%
  \BibitemOpen
  \bibfield  {author} {\bibinfo {author} {\bibfnamefont {A.}~\bibnamefont
  {Gottwald}}, \bibinfo {author} {\bibfnamefont {R.}~\bibnamefont {Klein}},
  \bibinfo {author} {\bibfnamefont {R.}~\bibnamefont {M\"uller}}, \bibinfo
  {author} {\bibfnamefont {M.}~\bibnamefont {Richter}}, \bibinfo {author}
  {\bibfnamefont {F.}~\bibnamefont {Scholze}}, \bibinfo {author} {\bibfnamefont
  {R.}~\bibnamefont {Thornagel}}, \ and\ \bibinfo {author} {\bibfnamefont
  {G.}~\bibnamefont {Ulm}},\ }\href
  {http://stacks.iop.org/0026-1394/49/i=2/a=S146} {\bibfield  {journal}
  {\bibinfo  {journal} {Metrologia}\ }\textbf {\bibinfo {volume} {49}},\
  \bibinfo {pages} {S146} (\bibinfo {year} {2012})}\BibitemShut {NoStop}%
\bibitem [{\citenamefont {Fink}\ \emph {et~al.}(1994)\citenamefont {Fink},
  \citenamefont {N\"ucker}, \citenamefont {Pellegrin}, \citenamefont {Romberg},
  \citenamefont {Alexander},\ and\ \citenamefont {Knupfer}}]{Fink1994}%
  \BibitemOpen
  \bibfield  {author} {\bibinfo {author} {\bibfnamefont {J.}~\bibnamefont
  {Fink}}, \bibinfo {author} {\bibfnamefont {N.}~\bibnamefont {N\"ucker}},
  \bibinfo {author} {\bibfnamefont {E.}~\bibnamefont {Pellegrin}}, \bibinfo
  {author} {\bibfnamefont {H.}~\bibnamefont {Romberg}}, \bibinfo {author}
  {\bibfnamefont {M.}~\bibnamefont {Alexander}}, \ and\ \bibinfo {author}
  {\bibfnamefont {M.}~\bibnamefont {Knupfer}},\ }\href {\doibase
  http://dx.doi.org/10.1016/0368-2048(93)01857-B} {\bibfield  {journal}
  {\bibinfo  {journal} {J. Electron Spectrosc. Relat. Phenom.}\ }\textbf
  {\bibinfo {volume} {66}},\ \bibinfo {pages} {395 } (\bibinfo {year}
  {1994})}\BibitemShut {NoStop}%
\bibitem [{\citenamefont {Fink}(1989)}]{Fink1989}%
  \BibitemOpen
  \bibfield  {author} {\bibinfo {author} {\bibfnamefont {J.}~\bibnamefont
  {Fink}}\ }(\bibinfo  {publisher} {Academic Press},\ \bibinfo {year} {1989})\
  pp.\ \bibinfo {pages} {121 -- 232}\BibitemShut {NoStop}%
\bibitem [{\citenamefont {Roth}\ \emph
  {et~al.}(2014{\natexlab{b}})\citenamefont {Roth}, \citenamefont {K\"onig},
  \citenamefont {Fink}, \citenamefont {B\"uchner},\ and\ \citenamefont
  {Knupfer}}]{Roth2014}%
  \BibitemOpen
  \bibfield  {author} {\bibinfo {author} {\bibfnamefont {F.}~\bibnamefont
  {Roth}}, \bibinfo {author} {\bibfnamefont {A.}~\bibnamefont {K\"onig}},
  \bibinfo {author} {\bibfnamefont {J.}~\bibnamefont {Fink}}, \bibinfo {author}
  {\bibfnamefont {B.}~\bibnamefont {B\"uchner}}, \ and\ \bibinfo {author}
  {\bibfnamefont {M.}~\bibnamefont {Knupfer}},\ }\href {\doibase
  http://dx.doi.org/10.1016/j.elspec.2014.05.007} {\bibfield  {journal}
  {\bibinfo  {journal} {J. Electron Spectrosc. Relat. Phenom.}\ }\textbf
  {\bibinfo {volume} {195}},\ \bibinfo {pages} {85 } (\bibinfo {year}
  {2014}{\natexlab{b}})}\BibitemShut {NoStop}%
\bibitem [{\citenamefont {Liao}\ and\ \citenamefont
  {Scheiner}(2001)}]{Liao2001}%
  \BibitemOpen
  \bibfield  {author} {\bibinfo {author} {\bibfnamefont {M.-S.}\ \bibnamefont
  {Liao}}\ and\ \bibinfo {author} {\bibfnamefont {S.}~\bibnamefont
  {Scheiner}},\ }\href {\doibase http://dx.doi.org/10.1063/1.1367374}
  {\bibfield  {journal} {\bibinfo  {journal} {J. Chem. Phys.}\ }\textbf
  {\bibinfo {volume} {114}},\ \bibinfo {pages} {9780} (\bibinfo {year}
  {2001})}\BibitemShut {NoStop}%
\bibitem [{\citenamefont {Ellis}\ \emph {et~al.}(2006)\citenamefont {Ellis},
  \citenamefont {Park}, \citenamefont {Ulrich}, \citenamefont {Hulbert},\ and\
  \citenamefont {Rowe}}]{Trinity2006}%
  \BibitemOpen
  \bibfield  {author} {\bibinfo {author} {\bibfnamefont {T.~S.}\ \bibnamefont
  {Ellis}}, \bibinfo {author} {\bibfnamefont {K.~T.}\ \bibnamefont {Park}},
  \bibinfo {author} {\bibfnamefont {M.~D.}\ \bibnamefont {Ulrich}}, \bibinfo
  {author} {\bibfnamefont {S.~L.}\ \bibnamefont {Hulbert}}, \ and\ \bibinfo
  {author} {\bibfnamefont {J.~E.}\ \bibnamefont {Rowe}},\ }\href {\doibase
  http://dx.doi.org/10.1063/1.2364034} {\bibfield  {journal} {\bibinfo
  {journal} {J. Appl. Phys.}\ }\textbf {\bibinfo {volume} {100}},\ \bibinfo
  {eid} {093515} (\bibinfo {year} {2006})}\BibitemShut {NoStop}%
\bibitem [{\citenamefont {Gargiani}\ \emph {et~al.}(2010)\citenamefont
  {Gargiani}, \citenamefont {Angelucci}, \citenamefont {Mariani},\ and\
  \citenamefont {Betti}}]{Gargiani2010}%
  \BibitemOpen
  \bibfield  {author} {\bibinfo {author} {\bibfnamefont {P.}~\bibnamefont
  {Gargiani}}, \bibinfo {author} {\bibfnamefont {M.}~\bibnamefont {Angelucci}},
  \bibinfo {author} {\bibfnamefont {C.}~\bibnamefont {Mariani}}, \ and\
  \bibinfo {author} {\bibfnamefont {M.~G.}\ \bibnamefont {Betti}},\ }\href
  {\doibase 10.1103/PhysRevB.81.085412} {\bibfield  {journal} {\bibinfo
  {journal} {Phys. Rev. B}\ }\textbf {\bibinfo {volume} {81}},\ \bibinfo
  {pages} {085412} (\bibinfo {year} {2010})}\BibitemShut {NoStop}%
\bibitem [{\citenamefont {Calzolari}\ \emph
  {et~al.}(2007{\natexlab{b}})\citenamefont {Calzolari}, \citenamefont
  {Ferretti},\ and\ \citenamefont {Nardelli}}]{Calzolari2007}%
  \BibitemOpen
  \bibfield  {author} {\bibinfo {author} {\bibfnamefont {A.}~\bibnamefont
  {Calzolari}}, \bibinfo {author} {\bibfnamefont {A.}~\bibnamefont {Ferretti}},
  \ and\ \bibinfo {author} {\bibfnamefont {M.~B.}\ \bibnamefont {Nardelli}},\
  }\href {http://stacks.iop.org/0957-4484/18/i=42/a=424013} {\bibfield
  {journal} {\bibinfo  {journal} {Nanotechnology}\ }\textbf {\bibinfo {volume}
  {18}},\ \bibinfo {pages} {424013} (\bibinfo {year}
  {2007}{\natexlab{b}})}\BibitemShut {NoStop}%
\bibitem [{\citenamefont {Weaver}(1992)}]{Weaver1992}%
  \BibitemOpen
  \bibfield  {author} {\bibinfo {author} {\bibfnamefont {J.}~\bibnamefont
  {Weaver}},\ }\href {\doibase http://dx.doi.org/10.1016/0022-3697(92)90237-8}
  {\bibfield  {journal} {\bibinfo  {journal} {J. Phys. Chem. Solids}\ }\textbf
  {\bibinfo {volume} {53}},\ \bibinfo {pages} {1433 } (\bibinfo {year}
  {1992})}\BibitemShut {NoStop}%
\bibitem [{\citenamefont {Golden}\ \emph {et~al.}(1995)\citenamefont {Golden},
  \citenamefont {Knupfer}, \citenamefont {Fink}, \citenamefont {Armbruster},
  \citenamefont {Cummins}, \citenamefont {Romberg}, \citenamefont {Roth},
  \citenamefont {Sing}, \citenamefont {Schmidt},\ and\ \citenamefont
  {Sohmen}}]{Golden1995}%
  \BibitemOpen
  \bibfield  {author} {\bibinfo {author} {\bibfnamefont {M.~S.}\ \bibnamefont
  {Golden}}, \bibinfo {author} {\bibfnamefont {M.}~\bibnamefont {Knupfer}},
  \bibinfo {author} {\bibfnamefont {J.}~\bibnamefont {Fink}}, \bibinfo {author}
  {\bibfnamefont {J.~F.}\ \bibnamefont {Armbruster}}, \bibinfo {author}
  {\bibfnamefont {T.~R.}\ \bibnamefont {Cummins}}, \bibinfo {author}
  {\bibfnamefont {H.~A.}\ \bibnamefont {Romberg}}, \bibinfo {author}
  {\bibfnamefont {M.}~\bibnamefont {Roth}}, \bibinfo {author} {\bibfnamefont
  {M.}~\bibnamefont {Sing}}, \bibinfo {author} {\bibfnamefont {M.}~\bibnamefont
  {Schmidt}}, \ and\ \bibinfo {author} {\bibfnamefont {E.}~\bibnamefont
  {Sohmen}},\ }\href {http://stacks.iop.org/0953-8984/7/i=43/a=004} {\bibfield
  {journal} {\bibinfo  {journal} {J. Phys.: Condens. Matter}\ }\textbf
  {\bibinfo {volume} {7}},\ \bibinfo {pages} {8219} (\bibinfo {year}
  {1995})}\BibitemShut {NoStop}%
\bibitem [{\citenamefont {Koch}(2007)}]{Koch2007}%
  \BibitemOpen
  \bibfield  {author} {\bibinfo {author} {\bibfnamefont {N.}~\bibnamefont
  {Koch}},\ }\href {\doibase 10.1002/cphc.200700177} {\bibfield  {journal}
  {\bibinfo  {journal} {Chem. Phys. Chem.}\ }\textbf {\bibinfo {volume} {8}},\
  \bibinfo {pages} {1438} (\bibinfo {year} {2007})}\BibitemShut {NoStop}%
\bibitem [{\citenamefont {V\'{a}zquez}\ \emph {et~al.}(2007)\citenamefont
  {V\'{a}zquez}, \citenamefont {Dappe}, \citenamefont {Ortega},\ and\
  \citenamefont {Flores}}]{Vazquez2007}%
  \BibitemOpen
  \bibfield  {author} {\bibinfo {author} {\bibfnamefont {H.}~\bibnamefont
  {V\'{a}zquez}}, \bibinfo {author} {\bibfnamefont {Y.}~\bibnamefont {Dappe}},
  \bibinfo {author} {\bibfnamefont {J.}~\bibnamefont {Ortega}}, \ and\ \bibinfo
  {author} {\bibfnamefont {F.}~\bibnamefont {Flores}},\ }\href {\doibase
  http://dx.doi.org/10.1016/j.apsusc.2007.07.047} {\bibfield  {journal}
  {\bibinfo  {journal} {Appl. Surf. Sci.}\ }\textbf {\bibinfo {volume} {254}},\
  \bibinfo {pages} {378 } (\bibinfo {year} {2007})}\BibitemShut {NoStop}%
\bibitem [{\citenamefont {Flores}\ \emph {et~al.}(2009)\citenamefont {Flores},
  \citenamefont {Ortega},\ and\ \citenamefont {Vazquez}}]{Flores2009}%
  \BibitemOpen
  \bibfield  {author} {\bibinfo {author} {\bibfnamefont {F.}~\bibnamefont
  {Flores}}, \bibinfo {author} {\bibfnamefont {J.}~\bibnamefont {Ortega}}, \
  and\ \bibinfo {author} {\bibfnamefont {H.}~\bibnamefont {Vazquez}},\ }\href
  {http://dx.doi.org/10.1039/B902492C} {\bibfield  {journal} {\bibinfo
  {journal} {Phys. Chem. Chem. Phys.}\ }\textbf {\bibinfo {volume} {11}},\
  \bibinfo {pages} {8658} (\bibinfo {year} {2009})}\BibitemShut {NoStop}%
\bibitem [{\citenamefont {Oehzelt}\ \emph {et~al.}(2014)\citenamefont
  {Oehzelt}, \citenamefont {Koch},\ and\ \citenamefont {Heimel}}]{Oehzelt2014}%
  \BibitemOpen
  \bibfield  {author} {\bibinfo {author} {\bibfnamefont {M.}~\bibnamefont
  {Oehzelt}}, \bibinfo {author} {\bibfnamefont {N.}~\bibnamefont {Koch}}, \
  and\ \bibinfo {author} {\bibfnamefont {G.}~\bibnamefont {Heimel}},\
  }\href@noop {} {\bibfield  {journal} {\bibinfo  {journal} {Nat Commun.}\
  }\textbf {\bibinfo {volume} {5}},\ \bibinfo {pages} {4174} (\bibinfo {year}
  {2014})}\BibitemShut {NoStop}%
\bibitem [{\citenamefont {Molodtsova}\ and\ \citenamefont
  {Knupfer}(2006)}]{Molodtsova2006}%
  \BibitemOpen
  \bibfield  {author} {\bibinfo {author} {\bibfnamefont {O.~V.}\ \bibnamefont
  {Molodtsova}}\ and\ \bibinfo {author} {\bibfnamefont {M.}~\bibnamefont
  {Knupfer}},\ }\href {\doibase http://dx.doi.org/10.1063/1.2175468} {\bibfield
   {journal} {\bibinfo  {journal} {J. Appl. Phys.}\ }\textbf {\bibinfo {volume}
  {99}},\ \bibinfo {pages} {053704} (\bibinfo {year} {2006})}\BibitemShut
  {NoStop}%
\bibitem [{\citenamefont {Sohmen}\ \emph {et~al.}(1992)\citenamefont {Sohmen},
  \citenamefont {Fink},\ and\ \citenamefont {Kr\"atschmer}}]{Sohmen1992}%
  \BibitemOpen
  \bibfield  {author} {\bibinfo {author} {\bibfnamefont {E.}~\bibnamefont
  {Sohmen}}, \bibinfo {author} {\bibfnamefont {J.}~\bibnamefont {Fink}}, \ and\
  \bibinfo {author} {\bibfnamefont {W.}~\bibnamefont {Kr\"atschmer}},\ }\href
  {http://dx.doi.org/10.1007/BF01323552} {\bibfield  {journal} {\bibinfo
  {journal} {Z. Phys. B Condensed Matter}\ }\textbf {\bibinfo {volume} {86}},\
  \bibinfo {pages} {87} (\bibinfo {year} {1992})}\BibitemShut {NoStop}%
\bibitem [{\citenamefont {Hartmann}\ \emph {et~al.}(1995)\citenamefont
  {Hartmann}, \citenamefont {Zigone}, \citenamefont {Martinez}, \citenamefont
  {Shirley}, \citenamefont {Benedict}, \citenamefont {Louie}, \citenamefont
  {Fuhrer},\ and\ \citenamefont {Zettl}}]{Hartmann1995}%
  \BibitemOpen
  \bibfield  {author} {\bibinfo {author} {\bibfnamefont {C.}~\bibnamefont
  {Hartmann}}, \bibinfo {author} {\bibfnamefont {M.}~\bibnamefont {Zigone}},
  \bibinfo {author} {\bibfnamefont {G.}~\bibnamefont {Martinez}}, \bibinfo
  {author} {\bibfnamefont {E.~L.}\ \bibnamefont {Shirley}}, \bibinfo {author}
  {\bibfnamefont {L.~X.}\ \bibnamefont {Benedict}}, \bibinfo {author}
  {\bibfnamefont {S.~G.}\ \bibnamefont {Louie}}, \bibinfo {author}
  {\bibfnamefont {M.~S.}\ \bibnamefont {Fuhrer}}, \ and\ \bibinfo {author}
  {\bibfnamefont {A.}~\bibnamefont {Zettl}},\ }\href@noop {} {\bibfield
  {journal} {\bibinfo  {journal} {Phys. Rev. B}\ }\textbf {\bibinfo {volume}
  {52}},\ \bibinfo {pages} {R5550} (\bibinfo {year} {1995})}\BibitemShut
  {NoStop}%
\bibitem [{\citenamefont {Knupfer}\ and\ \citenamefont
  {Fink}(1999)}]{Knupfer1999}%
  \BibitemOpen
  \bibfield  {author} {\bibinfo {author} {\bibfnamefont {M.}~\bibnamefont
  {Knupfer}}\ and\ \bibinfo {author} {\bibfnamefont {J.}~\bibnamefont {Fink}},\
  }\href@noop {} {\bibfield  {journal} {\bibinfo  {journal} {Phys. Rev. B}\
  }\textbf {\bibinfo {volume} {60}},\ \bibinfo {pages} {10731} (\bibinfo {year}
  {1999})}\BibitemShut {NoStop}%
\bibitem [{\citenamefont {Fielding}\ and\ \citenamefont
  {Mackay}(1964)}]{Fielding1964}%
  \BibitemOpen
  \bibfield  {author} {\bibinfo {author} {\bibfnamefont {P.}~\bibnamefont
  {Fielding}}\ and\ \bibinfo {author} {\bibfnamefont {A.}~\bibnamefont
  {Mackay}},\ }\href@noop {} {\bibfield  {journal} {\bibinfo  {journal} {Aust.
  J. Chem.}\ }\textbf {\bibinfo {volume} {17}},\ \bibinfo {pages} {750}
  (\bibinfo {year} {1964})}\BibitemShut {NoStop}%
\bibitem [{\citenamefont {G.~Engelsma}(1962)}]{Engelsma1962}%
  \BibitemOpen
  \bibfield  {author} {\bibinfo {author} {\bibfnamefont {E.~M. M.~C.}\
  \bibnamefont {G.~Engelsma}, \bibfnamefont {A.~Yamamoto}},\ }\href@noop {}
  {\bibfield  {journal} {\bibinfo  {journal} {J. Phys. Chem.}\ }\textbf
  {\bibinfo {volume} {66}},\ \bibinfo {pages} {2517} (\bibinfo {year}
  {1962})}\BibitemShut {NoStop}%
\bibitem [{\citenamefont {Lever}\ \emph {et~al.}(1981)\citenamefont {Lever},
  \citenamefont {Pickens}, \citenamefont {Minor}, \citenamefont {Licoccia},
  \citenamefont {Ramaswamy},\ and\ \citenamefont {Magnell}}]{Lever1981}%
  \BibitemOpen
  \bibfield  {author} {\bibinfo {author} {\bibfnamefont {A.~B.~P.}\
  \bibnamefont {Lever}}, \bibinfo {author} {\bibfnamefont {S.~R.}\ \bibnamefont
  {Pickens}}, \bibinfo {author} {\bibfnamefont {P.~C.}\ \bibnamefont {Minor}},
  \bibinfo {author} {\bibfnamefont {S.}~\bibnamefont {Licoccia}}, \bibinfo
  {author} {\bibfnamefont {B.~S.}\ \bibnamefont {Ramaswamy}}, \ and\ \bibinfo
  {author} {\bibfnamefont {K.}~\bibnamefont {Magnell}},\ }\href@noop {}
  {\bibfield  {journal} {\bibinfo  {journal} {JACS}\ }\textbf {\bibinfo
  {volume} {103}},\ \bibinfo {pages} {6800} (\bibinfo {year}
  {1981})}\BibitemShut {NoStop}%
\bibitem [{\citenamefont {Sharp}\ and\ \citenamefont
  {Abkowitz}(1973)}]{Sharp1973}%
  \BibitemOpen
  \bibfield  {author} {\bibinfo {author} {\bibfnamefont {J.~H.}\ \bibnamefont
  {Sharp}}\ and\ \bibinfo {author} {\bibfnamefont {M.}~\bibnamefont
  {Abkowitz}},\ }\href@noop {} {\bibfield  {journal} {\bibinfo  {journal} {J.
  Phys. Chem.}\ }\textbf {\bibinfo {volume} {77}},\ \bibinfo {pages} {477}
  (\bibinfo {year} {1973})}\BibitemShut {NoStop}%
\bibitem [{\citenamefont {Yoshida}\ \emph {et~al.}(1986)\citenamefont
  {Yoshida}, \citenamefont {Tokura},\ and\ \citenamefont {Koda}}]{Yoshida1986}%
  \BibitemOpen
  \bibfield  {author} {\bibinfo {author} {\bibfnamefont {H.}~\bibnamefont
  {Yoshida}}, \bibinfo {author} {\bibfnamefont {Y.}~\bibnamefont {Tokura}}, \
  and\ \bibinfo {author} {\bibfnamefont {T.}~\bibnamefont {Koda}},\ }\href
  {\doibase http://dx.doi.org/10.1016/0301-0104(86)87066-5} {\bibfield
  {journal} {\bibinfo  {journal} {Chem. Phys.}\ }\textbf {\bibinfo {volume}
  {109}},\ \bibinfo {pages} {375 } (\bibinfo {year} {1986})}\BibitemShut
  {NoStop}%
\bibitem [{\citenamefont {Auerhammer}\ \emph {et~al.}(2002)\citenamefont
  {Auerhammer}, \citenamefont {Knupfer}, \citenamefont {Peisert},\ and\
  \citenamefont {Fink}}]{Auerhammer2002}%
  \BibitemOpen
  \bibfield  {author} {\bibinfo {author} {\bibfnamefont {J.}~\bibnamefont
  {Auerhammer}}, \bibinfo {author} {\bibfnamefont {M.}~\bibnamefont {Knupfer}},
  \bibinfo {author} {\bibfnamefont {H.}~\bibnamefont {Peisert}}, \ and\
  \bibinfo {author} {\bibfnamefont {J.}~\bibnamefont {Fink}},\ }\href {\doibase
  http://dx.doi.org/10.1016/S0039-6028(02)01517-0} {\bibfield  {journal}
  {\bibinfo  {journal} {Surf. Sci.}\ }\textbf {\bibinfo {volume} {506}},\
  \bibinfo {pages} {333 } (\bibinfo {year} {2002})}\BibitemShut {NoStop}%
\bibitem [{\citenamefont {Gunaratne}\ \emph {et~al.}(2004)\citenamefont
  {Gunaratne}, \citenamefont {Kennedy}, \citenamefont {Kenney},\ and\
  \citenamefont {Rodgers}}]{Gunaratne2004}%
  \BibitemOpen
  \bibfield  {author} {\bibinfo {author} {\bibfnamefont {T.}~\bibnamefont
  {Gunaratne}}, \bibinfo {author} {\bibfnamefont {V.~O.}\ \bibnamefont
  {Kennedy}}, \bibinfo {author} {\bibfnamefont {M.~E.}\ \bibnamefont {Kenney}},
  \ and\ \bibinfo {author} {\bibfnamefont {M.~A.~J.}\ \bibnamefont {Rodgers}},\
  }\href@noop {} {\bibfield  {journal} {\bibinfo  {journal} {J. Phys. Chem. A}\
  }\textbf {\bibinfo {volume} {108}},\ \bibinfo {pages} {2576} (\bibinfo {year}
  {2004})}\BibitemShut {NoStop}%
\bibitem [{\citenamefont {Knupfer}\ \emph {et~al.}(2004)\citenamefont
  {Knupfer}, \citenamefont {Schwieger}, \citenamefont {Peisert},\ and\
  \citenamefont {Fink}}]{Knupfer2004}%
  \BibitemOpen
  \bibfield  {author} {\bibinfo {author} {\bibfnamefont {M.}~\bibnamefont
  {Knupfer}}, \bibinfo {author} {\bibfnamefont {T.}~\bibnamefont {Schwieger}},
  \bibinfo {author} {\bibfnamefont {H.}~\bibnamefont {Peisert}}, \ and\
  \bibinfo {author} {\bibfnamefont {J.}~\bibnamefont {Fink}},\ }\href@noop {}
  {\bibfield  {journal} {\bibinfo  {journal} {Phys. Rev. B}\ }\textbf {\bibinfo
  {volume} {69}},\ \bibinfo {pages} {165210} (\bibinfo {year}
  {2004})}\BibitemShut {NoStop}%
\bibitem [{\citenamefont {Rand}\ \emph {et~al.}(2005)\citenamefont {Rand},
  \citenamefont {Xue}, \citenamefont {Uchida},\ and\ \citenamefont
  {Forrest}}]{Rand2005}%
  \BibitemOpen
  \bibfield  {author} {\bibinfo {author} {\bibfnamefont {B.~P.}\ \bibnamefont
  {Rand}}, \bibinfo {author} {\bibfnamefont {J.}~\bibnamefont {Xue}}, \bibinfo
  {author} {\bibfnamefont {S.}~\bibnamefont {Uchida}}, \ and\ \bibinfo {author}
  {\bibfnamefont {S.~R.}\ \bibnamefont {Forrest}},\ }\href@noop {} {\bibfield
  {journal} {\bibinfo  {journal} {J. Appl. Phys.}\ }\textbf {\bibinfo {volume}
  {98}} (\bibinfo {year} {2005})}\BibitemShut {NoStop}%
\end{thebibliography}
\end{document}